\documentclass[12pt,a4paper]{article}
\usepackage[latin1]{inputenc}
\usepackage{a4wide}
\usepackage{amssymb}
\usepackage{ifpdf}
\ifpdf
\usepackage[pdftex]{graphicx}
\usepackage[pdftex,unicode,implicit]{hyperref}
\hypersetup{%
  pdftitle    = {Characterization of all
 the supersymmetric solutions of N=1,d=5 gauged supergravity},
  pdfkeywords = {Supersymmetry, gauged supergravity, real special geometry, 
                 BPS solutions},
  pdfauthor   = {J. Bellorín, T. Ortín},
  pdfcreator  = {pdf\LaTeXe\ with package \flqq hyperref\frqq},
  pdfproducer = {pdf\LaTeXe\ with package \flqq hyperref\frqq},
  pdfpagemode = None,  
  pdffitwindow= true,  
  unicode     = true,
  plainpages  = true,
  colorlinks  = true,  
  citecolor   = blue,
  urlcolor    = red,
  linkcolor   = black
}
\newcommand{\hepth}[1]{arXiv:{\tt 
\href{http://www.arXiv.org/abs/hep-th/#1}{hep-th/#1}}}

\newcommand{\arxiv}[1]{{\tt 
\href{http://www.arXiv.org/abs/#1}{arXiv:#1}}}
\else
  \usepackage[dvips]{graphicx}
  \usepackage[unicode,implicit]{hyperref}
  \newcommand{\hepth}[1]{arXiv:{\tt hep-th/#1}}

  \newcommand{\arxiv}[1]{{\tt arXiv:#1}}
\fi
\makeatletter
\@addtoreset{equation}{section}
\makeatother

\begin{document}

\begin{flushright}
\small
IFT-UAM/CSIC-07-22\\
{\bf arXiv:0705.2567}\\
May $17^{\rm th}$, $2007$
\normalsize
\end{flushright}
\begin{center}
\vspace{2cm}
{\LARGE {\bf Characterization of all the supersymmetric solutions of\\[.5cm]
    gauged $N=1,d=5$ supergravity}} 
\vspace{2cm}

{\sl\large Jorge Bellor\'{\i}n}
\footnote{E-mail: {\tt Jorge.Bellorin@uam.es}}
{\sl\large and Tom{\'a}s Ort\'{\i}n}
\footnote{E-mail: {\tt Tomas.Ortin@cern.ch}}

\vspace{1cm}

{\it Instituto de F\'{\i}sica Te\'orica UAM/CSIC\\
Facultad de Ciencias C-XVI,  C.U.~Cantoblanco,  E-28049-Madrid, Spain}\\

\vspace{2cm}


{\bf Abstract}

\end{center}

\begin{quotation}\small
  We find a complete characterization of all the supersymmetric solutions of
  non-Abelian gauged $N=1,d=5$ supergravity coupled to vector multiplets and
  hypermultiplets: the generic forms of the metrics as functions of the
  scalars and vector fields plus the equations that all these must satisfy.
  These equations are now a complicated non-linear system and there it seems
  impossible to produce an algorithm to construct systematically all
  supersymmetric solutions.

\end{quotation}

\newpage

\pagestyle{plain}


\tableofcontents

\vspace{1cm}

\section{Introduction}

Supersymmetric classical solutions of supergravity theories have played and
continue playing a crucial role in many important developments such as
$AdS/CFT$ correspondence, stringy black-hole Physics etc. This is why a great
effort has been made to find , classify, or, at least, characterize all of
them. 

This program has been carried out to completion in several lower-dimensional
theories. The first work of this kind was carried in pure, ungauged, $N=2,d=4$
supergravity by Tod in his pioneering 1983 work \cite{Tod:1983pm} and it has
been extended to the gauged case in Ref.~\cite{Caldarelli:2003pb} and to
include the coupling to general (ungauged) vector multiplets and
hypermultiplets in Refs.~\cite{Meessen:2006tu} and \cite{Huebscher:2006mr},
respectively. Pure $N=4,d=4$ supergravity was dealt with in
Refs.~\cite{Tod:1995jf,Bellorin:2005zc}. The minimal $N=1,d=5$ theory was
worked out in Ref.~\cite{Gauntlett:2002nw} and the results were extended to
the gauged case in Ref.~\cite{Gauntlett:2003fk}, and the coupling to an
arbitrary number of vector multiplets and their Abelian gaugings was
considered in Refs.~\cite{Gutowski:2004yv,Gutowski:2005id}\footnote{Previous
  work on these theories can be found in
  Refs.~\cite{Chamseddine:1998yv,Sabra:1997yd}.}.  The inclusion of (ungauged)
hypermultiplets was considered in \cite{Bellorin:2006yr}\footnote{Previous
  partial results on that problem were presented in
  Refs.~\cite{Cacciatori:2002qx,Celi:2003qk,Cacciatori:2004qm}.}  and the goal
of this paper is to further extend these results to include non-Abelian
gaugings.

The minimal $d=6$ SUGRA was dealt with in
Refs.~\cite{Gutowski:2003rg,Chamseddine:2003yy}, some gaugings were considered
in Ref.~\cite{Cariglia:2004kk} and the coupling to hypermultiplets has been
fully solved in Ref.~\cite{Jong:2006za}. 

All the works cited are essentially based on the method pioneered by Tod and
generalized by Gauntlett \textit{et al.} in
Ref.~\cite{Gauntlett:2002nw}\footnote{Further works based on the alternative
  methods of spinorial geometry are
  Refs.~\cite{Gauntlett:2002fz,Cacciatori:2007vn}.}. This method consists on
assuming the existence of one Killing spinor and then deriving consistency
conditions for this to be true. These conditions can be conveniently computed
on tensors constructed as bilinears of the Killing spinors and constrain the
form of the fields of the supersymmetric configuration. Finally the equations
of motion have to be imposed on the constrained configurations, leading to
simpler equations involving the undetermined components of the fields. This
is the method that we are going to use here.

In the simplest cases (ungauged supergravities coupled to vector multiplets)
the equations that have to be solved are uncoupled, typically linear, and can
be solved in a systematic way. We can then construct supersymmetric solutions
for those theories in a systematic way. The coupling to hypermultiplets
\cite{Huebscher:2006mr,Bellorin:2006yr,Jong:2006za} introduces new equations
which, not only are non-linear but are coupled and have to be solved
simultaneously. In particular one finds supersymmetry implies that the
hyperscalar functions have to solve a nonlinear equation and, at the same
time, they must be such that the pullback of the quaternionic $SU(2)$
connection is gauge equivalent to the anti-selfdual part of the spin
connection of the base space. Finding base spaces and hyperscalars that
satisfy these two conditions is highly non-trivial and it is not known how to
do it systematically. Still, once those two conditions are solved, the
remaining equations are linear and uncoupled (Laplace equations for
independent functions on the base space).

As we are going to see, the introduction of non-Abelian gaugings leads to yet
more non-linear and coupled equations. This was to be expected since, for
instance, the requirement of having unbroken supersymmetry in Euclidean $d=4$
super-Yang-Mills theories still leaves us with non-linear equations to be
solved, namely finding gauge potentials that give self- or anti-self-dual
field strengths. In the case that we are going to study, timelike
supersymmetry implies that the hyperscalar functions have to solve a nonlinear
equation which involves, not only the hyperscalars, but the gauge potentials
and the scalars belonging to the vector multiplets which, at the same time,
must satisfy other equations. Simultaneously, the hyperscalar functions must be
such that the covariant pullback of the quaternionic $SU(2)$ connection is
gauge equivalent to the anti-selfdual part of the spin connection of the base
space. This is another condition that involves the hyperscalars, the gauge
connection and the base space metric.

Our results are, thus, less satisfactory than in the simplest cases, even if
they are complete characterizations of the necessary and sufficient conditions
for any field configuration to be a supersymmetric solution. Constructing
supersymmetric solutions of these theories is a difficult problem even though
we know the minimal set of equations that should be solved\footnote{A
  solutions could be immediately constructed, though, by dimensionally
  reducing the 6-dimensional dyonic string of Ref.~\cite{Jong:2006za}.}. We,
thus, leave for future work the construction of particular examples
\cite{kn:BCMO}.


This paper is organized as follows: in Section~\ref{sec-n1d5mg} we present the
fields, Lagrangian and supersymmetry transformation rules of the theories that
we are going to study. In Section~\ref{sec-susy} we study the necessary and
sufficient conditions for a configuration to be supersymmetric. As usual, we
study separately the case in which the Killing vector constructed as bilinear
of the Killing spinor is timelike (Section~\ref{sec-timelike}) and null
(Section~\ref{sec-null}). In Section~\ref{sec-conclusions} we present our
conclusions. The appendix contains some useful formulae used in the main text
concerning the gauging of isometries and the definition and meaning of the
momentum map.


\section{$N=1,d=5$ supergravity with gaugings}
\label{sec-n1d5mg}

In this section we are going to briefly describe the action, equations of
motion and supersymmetry transformation rules of gauged $N=1,d=5$
supergravities\footnote{Gauging of $N=1,d=5$ supergravity theories was first
  considered in Ref.~\cite{Gunaydin:1983bi,Gunaydin:1984ak}. More general
  gaugings in the vector multiplet sector plus the tensor multiplet sector
  (which we are not considering here) were considered in
  Refs.~\cite{Gunaydin:1999zx,Gunaydin:2000ph} and hypermultiplets and their
  gaugings were considered in Ref.~\cite{Ceresole:2000jd}. More general
  gaugings of the tensor multiplets and $N=1,d=5$ supergravities that do not
  admit actions were considered in Ref.~\cite{Bergshoeff:2002qk}.}, which we
take from Ref.~\cite{Bergshoeff:2004kh}, relying in the description of the
ungauged theories given in Ref.~\cite{Bellorin:2006yr}, whose conventions we
follow.  Appendix~\ref{sec-gauging} contains a description of the gauging of
the isometries of the scalar manifolds of the theory in which the definitions
of the covariant derivatives $\mathfrak{D}$, gauge transformations and
momentum map $\vec{P}_{I}$ can be found.

The bosonic action of $N=1,d=5$ gauged supergravity is given by

\begin{equation}
\begin{array}{rcl}
S & = &  {\displaystyle\int} d^{5}x\sqrt{g}\
\biggl\{
R
+{\textstyle\frac{1}{2}}g_{xy}\mathfrak{D}_{\mu}\phi^{x}
\mathfrak{D}^{\mu}\phi^{y}
+{\textstyle\frac{1}{2}}g_{XY}\mathfrak{D}_{\mu} q^{X} 
\mathfrak{D}^{\mu} q^{Y} 
+\mathcal{V}(\phi,q)
-{\textstyle\frac{1}{4}} a_{IJ} F^{I\, \mu\nu}F^{J}{}_{\mu\nu}
\\ \\ & & 
\hspace{2cm}
+{\textstyle\frac{1}{12\sqrt{3}}}C_{IJK}
{\displaystyle\frac{\varepsilon^{\mu\nu\rho\sigma\alpha}}{\sqrt{g}}}
\left(
F^{I}{}_{\mu\nu}F^{J}{}_{\rho\sigma}A^{K}{}_{\alpha}
-{\textstyle\frac{1}{2}}gf_{LM}{}^{I} F^{J}{}_{\mu\nu} 
A^{K}{}_{\rho} A^{L}{}_{\sigma} A^{M}{}_{\alpha}
\right.
\\ \\ & & 
\hspace*{2cm}
\left.
+{\textstyle\frac{1}{10}} g^2 f_{LM}{}^{I} f_{NP}{}^{J} 
A^{K}{}_{\mu} A^{L}{}_{\nu} A^{M}{}_{\rho} A^{N}{}_{\sigma} A^{P}{}_{\alpha}
\right)
\biggr\}\,,
\end{array}
\end{equation}

\noindent
where 

\begin{equation}
\mathcal{V}(\phi,q) = g^{2} \left(
4C_{IJK}h^{I}\vec{P}^{J}\cdot\vec{P}^{K}
-{\textstyle\frac{3}{2}}h^{I}h^{J} k_{I}{}^{X} k_{J}{}^{Y} g_{XY} 
\right)\, ,
\end{equation}

\noindent
is the potential for the scalars.  In the limit of pure supergravity,
$n_{H}=n_{V}=0$, $\mathcal{V}$ becomes a cosmological constant.

The equations of motion, for which we use the same notation as in
Ref.~\cite{Bellorin:2006yr}, are 

\begin{eqnarray}
\mathcal{E}_{\mu\nu} 
& = &   
G_{\mu\nu}
-{\textstyle\frac{1}{2}}a_{IJ}\left(F^{I}{}_{\mu}{}^{\rho} F^{J}{}_{\nu\rho}
-{\textstyle\frac{1}{4}}g_{\mu\nu}F^{I\, \rho\sigma}F^{J}{}_{\rho\sigma}
\right)      
+{\textstyle\frac{1}{2}}g_{xy}\left(\mathfrak{D}_{\mu}\phi^{x} 
\mathfrak{D}_{\nu}\phi^{y}
-{\textstyle\frac{1}{2}}g_{\mu\nu}
\mathfrak{D}_\rho\phi^{x} \mathfrak{D}^{\rho}\phi^{y}\right)
\nonumber\\
& & \nonumber  \\
& & 
+{\textstyle\frac{1}{2}}g_{XY}\left(\mathfrak{D}_{\mu}q^{X}
\mathfrak{D}_{\nu}q^{Y}
-{\textstyle\frac{1}{2}}g_{\mu\nu} 
\mathfrak{D}_{\rho}q^{X}\mathfrak{D}^{\rho}q^{Y}
\right)  -{\textstyle\frac{1}{2}}g_{\mu\nu}\mathcal{V}\, ,
\label{eq:Emn} \\
& & \nonumber \\
g^{xy}\mathcal{E}_{y} 
& = & 
\mathfrak{D}_{\mu}\mathfrak{D}^{\mu}\phi^{x} 
+{\textstyle\frac{1}{4}} \partial^x
a_{IJ} F^{I\, \rho\sigma}F^{J}{}_{\rho\sigma}
-\partial^x\mathcal{V}
\label{eq:Ei} \\ & & \nonumber \\
g^{XY}\mathcal{E}_{Y} 
& = & 
\mathfrak{D}_{\mu}\mathfrak{D}^{\mu}q^{X} 
-\partial^X\mathcal{V}
\, , \label{eq:EX}\\
& & \nonumber \\
\mathcal{E}_{I}{}^{\mu} 
& = & 
\mathfrak{D}_{\nu} F_{I}{}^{\nu\mu}
+{\textstyle\frac{1}{4\sqrt{3}}} 
\frac{\varepsilon^{\mu\nu\rho\sigma\alpha}}{\sqrt{g}}
C_{IJK} F^{J}{}_{\nu\rho}F^K{}_{\sigma\alpha}
+g\left(
k_{I\,x} \mathfrak{D}^\mu\phi^x + k_{I\,X} \mathfrak{D}^\mu q^{X}
\right) \, .
\label{eq:ERm}
\end{eqnarray}

The supersymmetry transformation rules for the fermionic fields, evaluated on
vanishing fermions, are

\begin{eqnarray}
\delta_{\epsilon}\psi^{i}_{\mu} 
& = & 
\mathfrak{D}_{\mu}\epsilon^{i}
-{\textstyle\frac{1}{8\sqrt{3}}}h_{I}F^{I\,\alpha\beta}
\left(\gamma_{\mu\alpha\beta}-4g_{\mu\alpha}\gamma_\beta\right)
\epsilon^{i}
+{\textstyle\frac{1}{2\sqrt3}} g h^{I} \gamma_\mu\epsilon^j P_{I\,j}{}^{i}\, , 
\\
& & \nonumber \\ 
\delta_{\epsilon}\lambda^{ix} 
& = &  
{\textstyle\frac{1}{2}}\left(\not\!\!\mathfrak{D}\phi^{x} 
-{\textstyle\frac{1}{2}}h^{x}_{I}\not\!F^{I}\right)\epsilon^{i}
+ g h_{I}^{x}\epsilon^{j} P^{I}{}_{j}{}^{i}\, ,\\
& & \nonumber \\
\delta_{\epsilon}\zeta^{A} 
& = & 
{\textstyle\frac{1}{2}}f_{X}{}^{iA}\left(\not\!\!\mathfrak{D} q^{X} 
+\sqrt{3} gh^{I} k_{I}{}^{X}\right)\epsilon_{i}\, .
\end{eqnarray}

The supersymmetry transformation rules of the bosonic fields are exactly the
same as in the ungauged case \cite{Bellorin:2006yr}.  This implies that the
form of the Killing spinor identities (KSIs) relating the bosonic equations of
motion that one can derive from them \cite{Kallosh:1993wx,Bellorin:2005hy}
have the same form as in the ungauged case, given in \cite{Bellorin:2006yr},
although the equations of motion are now those given above, which differ from
those of the ungauged case by $g$-dependent terms. 

Apart from the identities derived in Ref.~\cite{Bellorin:2006yr} we have found
that, in the null case, there are additional identities that were overlooked
in that reference. We will discuss them in Section~\ref{sec-null}.


\section{Supersymmetric configurations and solutions}
\label{sec-susy}

Following the standard procedure, we assume that the KSEs 

\begin{eqnarray}
\label{gravitinokse}
\mathfrak{D}_{\mu}\epsilon^{i}
-{\textstyle\frac{1}{8\sqrt{3}}}h_{I}F^{I\,\alpha\beta}
\left(\gamma_{\mu\alpha\beta}-4g_{\mu\alpha}\gamma_\beta\right)\epsilon^{i} 
+{\textstyle\frac{1}{2\sqrt3}} g \gamma_\mu\epsilon^j h^{I} P_{I\,j}{}^{i}
& = & 0\, , \\
& & \nonumber \\
\label{gauginokse} 
\left(\not\!\!\mathfrak{D}\phi^{x} 
-{\textstyle\frac{1}{2}}h^{x}_{I}\not\!F^{I}\right)\epsilon^{i} 
+ 2g \epsilon^{j} h_{I}^{x} P^{I}{}_{j}{}^{i}
& = & 0\, ,\\
& & \nonumber \\
\label{hyperinokse}
f_{X}{}^{iA}
\left(\not\!\!\mathfrak{D} q^{X} 
+ \sqrt{3} gh^{I} k_{I}{}^{X}\right)\epsilon_{i} & = & 0\, ,
\end{eqnarray}

\noindent
admit at least one solution $\epsilon^{i}$ and we start deriving from them the
equations satisfied by the tensor bilinears that can be constructed from the
Killing spinor: the scalar $f$, the vector $V$ (both $SU(2)$ singlets) and the
three 2-forms $\Phi^{r}$, which form an $SU(2)$-triplet.

The fact that the Killing spinor satisfies Eq.~(\ref{gravitinokse}) leads to
the following differential equations for the bilinears:

\begin{eqnarray}
\label{df}
df & = & {\textstyle\frac{1}{\sqrt{3}}}h_{I}i_{V}F^{I}\, , \\
& & \nonumber \\
\label{killingvector}                
\nabla_{(\mu}V_{\nu)} & = & 0\, ,\\
& & \nonumber \\
\label{dV}
dV & = & 
-{\textstyle\frac{2}{\sqrt{3}}}fh_{I}F^{I}
-{\textstyle\frac{1}{\sqrt{3}}}h_{I}\star\left(F^{I}\wedge V\right)
-{\textstyle\frac{2}{\sqrt{3}}}gh^{I}\vec{P}_I\cdot\vec{\Phi}
\, ,\\
& & \nonumber \\
\label{nablaPhi}
\mathfrak{D}_{\alpha}\vec{\Phi}_{\beta\gamma} & = &
-{\textstyle\frac{1}{\sqrt{3}}}h_{I}F^{I\,\rho\sigma}
\left(g_{\rho[\beta}\star \vec{\Phi}_{\gamma]\alpha\sigma}
-g_{\rho\alpha}\star\vec{\Phi}_{\beta\gamma\sigma} 
-{\textstyle\frac{1}{2}}g_{\alpha[\beta}\star
\vec{\Phi}_{\gamma]\rho\sigma}
\right)
\nonumber \\ \nonumber \\ & &
+{\textstyle\frac{1}{\sqrt{3}}}gh^{I}\left(
 \vec{P}_I \times (\star\vec{\Phi})_{\alpha\beta\gamma}
+2g_{\alpha[\beta} V_{\gamma]} \vec{P}_{I}
\right)
\, ,
\end{eqnarray}

\noindent
where 

\begin{equation}
\mathfrak{D}_{\alpha}\vec{\Phi}_{\beta\gamma}
=
\nabla_{\alpha}\vec{\Phi}_{\beta\gamma} 
+2\vec{B}_{\alpha} 
\times \vec{\Phi}_{\beta\gamma}\,.  
\end{equation}
The differential equation for $\Phi^{r}$~(\ref{nablaPhi}) implies
\begin{equation}
\label{eq:covariantconstancy}
d\Phi^{r}+2\varepsilon^{rst}B^{s}
\wedge\Phi^{t}=
\sqrt{3} gh^{I}\epsilon^{rst}P^{s}{}_{I}\:\star\Phi^{t}\,.
\end{equation}

The fact that the Killing spinor satisfies Eqs.~(\ref{gauginokse}) and
(\ref{hyperinokse}) leads to the following algebraic equations for the tensor
bilinears:

\begin{eqnarray}
\label{lvphi}
V^\mu\mathfrak{D}_\mu\phi^{x} & = & 0\, ,\\
& & \nonumber \\
\label{FtimesPhi}
h^{x}_{I}F^{I}_{\alpha\beta}\vec{\Phi}^{\alpha\beta} 
& = & 4gfh_{I}^{x} \vec{P}^{I}\, ,\\
& & \nonumber \\
\label{prelvq}
V^{\mu} \mathfrak{D}_{\mu} q^{X} & = & -\sqrt{3} gfh^{I}k_{I}{}^{X}\, , \\
& & \nonumber \\
\label{dphi}
f\mathfrak{D}_\mu\phi^{x} - h^{x}_{I} F^{I}{}_{\mu\nu} V^\nu
& = & 0\, ,
\\ 
& & \nonumber \\ 
\label{eq:Phidphi}  
\vec{\Phi}_{\mu\nu} \mathfrak{D}^{\nu}\phi^{x}
+{\textstyle\frac{1}{4}}\epsilon_{\mu\nu\alpha\beta\gamma} 
h^{x}_{I} F^{I\,\nu\alpha}\vec{\Phi}^{\beta\gamma}
& = & -2gh^{x}_{I}\vec{P}^{I} V_{\mu} \, ,\\
& & \nonumber \\ 
\label{preholomorphicq}
f\mathfrak{D}_{\mu} q^{X} +\Phi^{r}{}_{\mu}{}^{\nu}\mathfrak{D}_{\nu} 
q^{Y} J^{r}{}_{Y}{}^{X} & = & -\sqrt{3} gh^{I}k_{I}{}^{X} V_{\mu}\, .
\end{eqnarray}

We are now ready to extract consequences of these equations. To start with,
Eq.~(\ref{killingvector}) says that $V$ is an isometry of the space-time
metric.  It is convenient to partially fix the $G$ gauge using the condition

\begin{equation}
\label{gaugefixing}
i_{V} A^{I} + \sqrt{3} fh^{I}=0\, ,
\end{equation}

\noindent
since then Eqs.~(\ref{prelvq}) and (\ref{lvphi}) become just

\begin{equation}
\label{lvq}
\mathcal{L}_{V} q^{X} = \mathcal{L}_{V} \phi^{x} =0\, ,
\end{equation}

\noindent
after use of the explicit expression of the Killing vectors $k_{I}{}^{x}$
Eq.~(\ref{eq:kix}).  Then, in this gauge, the scalars $q^{X},\phi^{x}$ and $f$
are independent of the coordinate adapted to the isometry (see Eq.~(\ref{df}).

The Fierz identities relate the modulus of the vector bilinear $V^{\mu}$ to
the scalar bilinear $f$: $V^{2}=f^{2}$. This means that, as usual, $V^{\mu}$
can be timelike or null.  We now consider separately the timelike ($f\neq 0$)
and null ($f=0$) cases.


\subsection{The timelike case}
\label{sec-timelike}


\subsubsection{The equations for the bilinears}

By definition this is the case in which $V^{\mu}$ is timelike,
$V^{2}=f^{2}>0$.  Introducing an adapted time coordinate $t$: $V =
\partial_{t}$ the metric can be written in the same form as in the ungauged
case:

\begin{equation}
\label{conforma-stationary}
ds^{2} = f^{2}\left(dt+\omega\right)^{2}
-f^{-1}h_{\underline{m}\underline{n}} dx^{m}dx^{n}\, ,
\end{equation}

\noindent
with $\omega$ and $h_{\underline{m}\underline{n}}$ independent of time.  As we
mentioned in the previous section, in the (partially) fixed $G$-gauge
($A^{I}{}_{t} = -\sqrt{3} fh^{I}$) $f,\phi^{x}$ and $q^{X}$ are
also time-independent.

The spatial metric $h_{\underline{m}\underline{n}}$ is endowed with an
\emph{almost} quaternionic structure, $\Phi^{r}{}_{m}{}^{n}$. This is an
algebraic property that only depends on the Fierz identities.

The next step is to obtain the form of the supersymmetric vector field
strength from Eqs.~(\ref{df}), (\ref{dV}), (\ref{FtimesPhi}) and
(\ref{dphi}). In order to write the result it is convenient to split the gauge
potential $A^{I}$ into an electric part, which is determined by the partial
gauge fixing $A^{I}{}_{t} = -\sqrt{3} fh^{I}$ and a magnetic part
$\hat{A}^{I}$ with only spatial components

\begin{eqnarray}
\label{eq:A} 
A^{I} & = & -\sqrt{3} h^{I} e^{0} + \hat{A}^{I} \, , \\
& & \nonumber \\
A^{I}{}_{\underline{m}} & = &  
\hat{A}^{I}{}_{\underline{m}} -\sqrt{3} f h^{I} \omega_{\underline{m}}\, .
\end{eqnarray}

Observe that, unlike the spatial components $A^{I}{}_{\underline{m}}$, the
components $\hat{A}^{I}{}_{\underline{m}}$ are invariant under local shifts of
the time coordinate: $t\rightarrow t+\delta t(x)$, $\omega \rightarrow \omega
-d\delta t(x)$ which do not change the form of the metric and, in particular,
leave the 4-dimensional metric $h_{\underline{m}\underline{n}}$ invariant. It
is the correct 4-dimensional potential in the Kaluza-Klein sense.

In terms of the new variables $\hat{A}^{I}$ the field strengths are given by 

\begin{equation}
\label{F}
F^{I} = 
-\sqrt{3}\ \hat{\mathfrak{D}}(h^{I}e^{0})+\hat{F}^{I}\, ,
\end{equation}

\noindent
where $ \hat{\mathfrak{D}}$ is the 4-dimensional spatial covariant
derivative\footnote{Strictly speaking the action of a 4-dimensional spatial
  covariant derivative on $e^{0}$ which contains $dt$ is not well-defined. It
  is understood that $ \hat{\mathfrak{D}}(fdt) = \hat{\mathfrak{D}}f \wedge
  dt$.} with respect to $\hat{A}^{I}$ and $\hat{F}^{I}$ is the non-Abelian
field strength of $\hat{A}^{I}$ and it is related to $\omega$ and the scalars
by

\begin{eqnarray}
\label{remanentconstraint}
h_{I}\hat{F}^{I+} & = &  
{\textstyle\frac{2}{\sqrt{3}}} (fd\omega)^{+} \, , \\
& & \nonumber \\
\label{eq:theta-}
\hat{F}^{I-} & = & 
-2gf^{-1}C^{IJK}h_{J}\vec{P}_{K}\cdot\vec{\Phi}\, . 
\end{eqnarray}

\noindent
$\tilde{F}^{I+}$ is related to  the 2-forms called
$\Theta^{I}$ in the ungauged case
\cite{Gauntlett:2002nw,Gutowski:2005id,Bellorin:2006yr} by 

\begin{equation}
\Theta^{I} = -{\textstyle\frac{1}{\sqrt{3}}}\hat{F}^{I+}\, .
\label{eq:susyA}
\end{equation}

It is also convenient to introduce the spatial $SU(2)$ connection
$\hat{\vec{B}}$

\begin{eqnarray}
\label{eq:hatB}
\hat{\vec{B}} & \equiv & 
\vec{A}+{\textstyle\frac{1}{2}} g\hat{A}^{I}\vec{P}_{I}\, ,\\
& & \nonumber \\
\vec{B} & = & -{\textstyle\frac{\sqrt{3}}{2}}h^{I}\vec{P}_{I}e^{0}
+\hat{\vec{B}}\, ,
\end{eqnarray}

\noindent 
and extend the definition of $\hat{\mathfrak{D}}_{\underline{m}}$ as the
spatial $G$- and $SU(2)$-covariant derivative made from the hatted connections
$\hat{A}^{I}$ and $\hat{\vec{B}}$, including also the affine and spin
connections of the base spatial manifold.

The Eq.~(\ref{preholomorphicq}) is purely spatial in the timelike case and it
becomes, in 4-dimensional notation\footnote{From now on spatial flat indices
  refer to the 4-dimensional spatial metric $h_{\underline{mn}}$.}

\begin{equation}
\label{triholomorphic}
\hat{\mathfrak{D}}_{m} q^{X} 
=
\Phi^{r}{}_{m}{}^{n}\ \hat{\mathfrak{D}}_{n} q^{Y}\ J^{r}{}_{Y}{}^{X}\, .
\end{equation}

\noindent
We notice that this equation, even though it is written in terms of covariant
derivatives, imposes no integrability condition on the gauge connections. That
is, as equation for $q^{X}$ it has always \emph{local} solution for any given
vector fields $\hat{A}^{I}$.

Projecting this equation along the Killing vectors $k_{I}$ yields an important
relation,

\begin{equation}
 k_{I\, X} \hat{\mathfrak{D}}_{m} q^{X} 
= 
-2\vec{\Phi}_{m}{}^{n} \hat{\mathfrak{D}}_{n}\vec{P}_{I}\,.
\label{eq:kDq-PhiDP}
\end{equation}

\noindent 
This projection is the one which appears in the Maxwell
equations~(\ref{eq:ERm}).

Let us study the differential equations for the two-forms $\vec{\Phi}$. The
projection of Eq.~(\ref{eq:covariantconstancy}) along $V$ says that they are
time-independent in the gauge (\ref{gaugefixing}):

\begin{equation}
\partial_{t}\vec{\Phi}_{mn}=0\, .
\end{equation}

The components of Eq.~(\ref{nablaPhi}) can be explicitly evaluated using the
5-dimensional metric Eq.~(\ref{conforma-stationary}) and the expression for
the field strengths Eq.~(\ref{F}).  Only the spatial components of the
5-dimensional covariant derivative give new information:

\begin{equation}
\label{constantJ}
\hat{\mathfrak{D}}_{m}\vec{\Phi}_{np} =  0\, .
\label{eq:phicovconst}
\end{equation}

\noindent
This is a condition for the anti-self-dual part of the spin connection $\xi$
of the base spatial manifold. Indeed we can solve for $\xi^{-}$ in an
arbitrary frame and $SU(2)$ gauge:

\begin{equation}
\label{eq:embedding}
\xi^{-}{}_{mnp}  =
-\hat{\vec{B}}_{m} \cdot \vec{\Phi}_{np} 
-{\textstyle\frac{1}{4}}\partial_{m}\vec{\Phi}_{nq}\cdot\vec{\Phi}_{qp}\, ,
\end{equation}

\noindent
where we have used the (Fierz) identity

\begin{equation}
\vec{\Phi}_{mn}\cdot\vec{\Phi}_{pq} =
\delta_{mp}\delta_{nq} - \delta_{mq}\delta_{np} - \epsilon_{mnpq}\, .
\label{eq:phidotphi}
\end{equation}

\noindent
The meaning of relation~(\ref{eq:embedding}) becomes clearer in a frame and
$SU(2)$ gauge in which the $\vec{\Phi}$s are constant: the $SU(2)$ connection
$\hat{\vec{B}}$ is embedded into the anti-self-dual part of the spin
connection of the base manifold. The same happenend in the ungauged case
\cite{Bellorin:2006yr} and, again, this embedding requires the action of the
$SU(2)$ generators in the fundamental and spinorial representation on spinors
to be identical, i.e.

\begin{equation}
\epsilon^{j}\ i\vec{\sigma}_{j}{}^{i} \; =\; 
{\textstyle\frac{1}{4}}\vec{\mathtt{J}}_{mn}\gamma^{mn}\ \epsilon^{i}\, ,
\end{equation}

\noindent
and these conditions will appear as projectors 

\begin{equation}
\label{eq:Pirpm}
\Pi^{r\, \pm}{}_{i}{}^{j}\; =\; {\textstyle\frac{1}{2}}\left[\ \delta
\ \pm\ {\textstyle\frac{i}{4}}
\not\!\mathtt{J}^{(r)}\sigma^{(r)}\right]_{i}{}^{j} \, ,
\end{equation}

\noindent
acting on the Killing spinors.

It is interesting to study the integrability condition of
Eq.~(\ref{constantJ}), which is

\begin{equation}
\left[ {\textstyle\frac{1}{4}} R^{-}{}_{mnkl} \vec{\Phi}^{kl}
+ \vec{R}_{mn}(\hat{\vec{B}}) \right] \times \vec{\Phi}_{pq} = 0\, ,
\end{equation}

\noindent
where $\vec{R}_{mn}(\hat{\vec{B}})$ is the curvature of $\hat{\vec{B}}$, which
is given by

\begin{equation}
\vec{R}_{mn}(\hat{\vec{B}}) = 
\hat{\mathfrak{D}}_{m} q^{X} \hat{\mathfrak{D}}_{n} q^{Y} 
\vec{R}_{XY}(\vec{\omega})
+{\textstyle\frac{1}{2}} g\hat{F}^{I}_{mn} \vec{P}_{I}
=
-{\textstyle\frac{1}{4}}\hat{\mathfrak{D}}_{m} q^{X} 
\hat{\mathfrak{D}}_{n} q^{Y} \vec{J}_{XY}
+{\textstyle\frac{1}{2}} g\hat{F}^{I}{}_{mn} \vec{P}_{I}\, ,
\end{equation}

\noindent
hence the integrability condition yields

\begin{equation}
R^{-}{}_{mnkl} \vec{\Phi}^{kl}
-\hat{\mathfrak{D}}_{m} q^{X} 
\hat{\mathfrak{D}}_{n} q^{Y} \vec{J}_{XY}
+2g \hat{F}^{I}{}_{mn} \vec{P}_{I}
= 0 \,.
\end{equation}

\noindent
We stress that this condition is equivalent to Eq.~(\ref{eq:embedding}).

Now if we contract this expression with $\vec{\Phi}^{pn}$ we can compare
it with Eq.~(\ref{eq:triholomorphic}) and doing so we obtain an expression
involving the Ricci tensor of the spatial metric $h_{\underline{mn}}$

\begin{equation}
R_{mn}(h) =
-{\textstyle\frac{1}{2}} \hat{\mathfrak{D}}_{m} q^{X}
\hat{\mathfrak{D}}_{n}q^{Y}g_{XY}
+2g^2f^{-1}C^{IJK}h_{I} \vec{P}_{J}\cdot\vec{P}_{K} \delta_{mn} 
+g \hat{F}^{I+}{}_{mp}\vec{\Phi}_{pn}\cdot\vec{P}_{I} \, ,
\label{eq:Ricci}
\end{equation}

\noindent
where we have used again the identity~(\ref{eq:phidotphi}), and consequently
the Ricci scalar

\begin{equation}
\label{eq:Ricciscalar}
R(h) =
-{\textstyle\frac{1}{2}} \hat{\mathfrak{D}}_{m} q^{X}
\hat{\mathfrak{D}}_{m} q^{Y}g_{XY}
+8g^2f^{-1}C^{IJK}h_{I} \vec{P}_{J}\cdot\vec{P}_{K}\, .
\end{equation}

In the ungauged case the Eq.~(\ref{eq:Ricci}) says that the Ricci tensor of
the spatial metric $h_{\underline{mn}}$ is proportional to the induced metric

\begin{equation}
R_{mn}(h) =
-{\textstyle\frac{1}{2}} \partial_{m} q^{X} \partial_{n} q^{Y}g_{XY} \,.
\end{equation}

\noindent
On the other hand in the gauged case we can solve the
Eq.~(\ref{eq:Ricciscalar}) for $f$,

\begin{equation}
\label{eq:f}
f = 
(8g^{2} C^{IJK}h_{I}\vec{P}_{J}\cdot\vec{P}_{K})/ 
(R(h) +{\textstyle\frac{1}{2}} \hat{\mathfrak{D}}_{m} q^{X}
\hat{\mathfrak{D}}_{m} q^{Y} g_{XY})\, .
\end{equation}




\subsubsection{Solving the Killing spinor equations}

We are now going to prove that the necessary conditions for having unbroken
supersymmetry that we have derived in the previous section are also
sufficient. Thus, we are going to assume that we have a configuration with a
metric of the form Eq.~(\ref{conforma-stationary}), a non-Abelian gauge
potential of the form Eq.~(\ref{eq:A}) with a field strength of the form
Eq.~(\ref{F}) satisfying Eqs.~(\ref{remanentconstraint}) and
(\ref{eq:theta-}), and hyperscalars such that Eqs.~(\ref{triholomorphic})
and~(\ref{eq:embedding}) are satisfied.

Substituting these expressions in the KSE associated to the gaugino SUSY
transformation rule Eq.~(\ref{gauginokse}), and expressing all terms in
4-dimensional language we get

\begin{equation}
f^{1/2}\left(2\not\!\!\hat{\mathfrak{D}}\phi^{x}
-{\textstyle\frac{\sqrt{3}}{2}}f^{1/2}h^{x}_{I}\not\!\tilde{\Theta}^{I+}\right)
R^{-}\epsilon^{i}
+2gh^{x}_{I}\vec{P}^{I}\cdot\left(i\vec{\sigma}_{j}{}^{i}
-{\textstyle\frac{1}{4}}\not\!\vec{\Phi}\delta_{j}{}^{i}\right)\epsilon^{j}
=0\, .
\label{eq:solvedgauginokse}
\end{equation}

\noindent
where

\begin{equation}
R^{\pm} \equiv {\textstyle\frac{1}{2}} \left(1\pm\gamma^{0}\right)\, ,
\hspace{1.5cm}
\Pi^{r\pm}{}_{j}{}^{i} \equiv
{\textstyle\frac{1}{2}}\left(\ \delta
\ \pm\ {\textstyle\frac{i}{4}}
\not\!\Phi^{(r)}\sigma^{(r)}\right)_{j}{}^{i}\, .
\end{equation}

\noindent
The projections 

\begin{equation}
\label{eq:projections}
\vec{\Pi}^{+}{}_{j}{}^{i} \epsilon^{j} =0\, ,
\hspace{1.5cm}
R^{-}\epsilon^{i} =0\, ,
\end{equation}

\noindent
are sufficient to solve it. All of them are necessary in the general case but
in particular cases in which the coefficients of the projectors in the above
and following equations vanish, only some of them may be necessary. The
discussion is entirely analogous to that of the ungauged case
\cite{Bellorin:2006yr}.

Substituting now in Eq.~(\ref{hyperinokse}) we get

\begin{equation}
f_{X}{}^{iA}\{f^{1/2} \not\!\!\hat{\mathfrak{D}}q^{X} \epsilon_{i}
+2\sqrt{3}gh^{I}k_{I}{}^{X}f_{X}{}^{iA}R^{-}\}\epsilon_{i}=0\, .
\label{eq:solvedhyperinokse}
\end{equation}

\noindent
The last term vanishes with the second projection of
Eqs.~(\ref{eq:projections}). On the other hand, from
Eq.~(\ref{triholomorphic}) we can derive the identity

\begin{equation}
f_{X}{}^{iA}\not\!\!\hat{\mathfrak{D}}q^{X} R^{+}
=
-f_{X}{}^{jA}\not\!\!\hat{\mathfrak{D}}q^{X}
\sum_{r}\left(\Pi^{r+}-\Pi^{r-}\right)_{j}{}^{i}
\,.
\label{eq:dqprojected}
\end{equation}


\noindent
Acting on $\epsilon_{i}$ and imposing again the
projections~(\ref{eq:projections}) we see that it leads to 

\begin{equation}
f_{X}{}^{iA}\not\!\!\hat{\mathfrak{D}}q^{X}\epsilon_{i} 
=
-3f_{X}{}^{iA}\not\!\!\hat{\mathfrak{D}}q^{X}\epsilon_{i} 
\hspace*{7mm}\Rightarrow\hspace*{7mm}
f_{X}{}^{iA}\not\!\!\hat{\mathfrak{D}}q^{X}\epsilon_{i} = 0\, .
\end{equation}

\noindent
Hence the hyperino KSE~(\ref{eq:solvedhyperinokse}) is also solved.



is automatically satisfied by constant Killing spinors upon the use of the
projections Eqs.~(\ref{eq:projections}).

Finally, the spatial components of the same equation take, using
$R^{-}\epsilon^{i}=0$, the form

\begin{equation}
\nabla_{m}\eta^{i} + \eta^{j}C_{m\,j}{}^{i}=0\, ,
\hspace*{10mm}
\eta^{i} \equiv f^{-1/2}\epsilon^{i} \,.
\end{equation}

\noindent
Using the relation~(\ref{eq:embedding}) and the projections, it becomes






\begin{equation}
\partial_{m}\eta^{i} 
+{\textstyle\frac{1}{16}}\partial_{m}\!\not\!\Phi_{j}{}^{i}\eta^{j}=0\, ,
\end{equation}

\noindent
where $\Phi_{i}{}^{j} = i\vec{\sigma}_{i}{}^{j}\cdot\vec{\Phi}$.

The solution of this equation is given in terms of the path-ordered
exponential

\begin{eqnarray}
\eta^{i}(x,x_0) =
P\exp\left(
-{\textstyle\frac{1}{16}}\int\limits_{x_{0}}^{x} dx_{1}^{\underline{m}}
\partial_{\underline{m}}\!\not\!\Phi_{j}{}^{i}(x_{1})
\right)\eta_{0}^{j}\, ,
\end{eqnarray}

\noindent
where $\eta_{0}^{i}$ is a constant spinor, or in a frame and $SU(2)$ gauge
where $\vec{\Phi}$ is constant, it is just the constant spinor $\eta_{0}^{i}$.

The analysis of the amount of unbroken supersymmetry is identical to that of
the ungauged case \cite{Bellorin:2006yr}.


\subsubsection{Supersymmetric solutions}

As we discussed at the end of Section~\ref{sec-n1d5mg}, the KSIs of the gauged
theories have the same form as those of the ungauged ones, which are given in
Ref.~\cite{Bellorin:2006yr}. There it was proven that timelike supersymmetric
configurations solve all the equations of motions if they solve the Maxwell
equations.  We are now going to impose those equations on the supersymmetric
configurations. It is possible to show that the Bianchi identities imply
the spatial components of the Maxwell equations for supersymmetric
configurations using Eq.~(\ref{eq:kDq-PhiDP})

\begin{equation}
\mathcal{E}_{I}{}^{m} = 2C_{IJK}h^{J}(\star\mathfrak{D} F^{K})^{\,0m}\, .
\end{equation}

\noindent
Thus we only need to impose the time component of the Maxwell equations on
the supersymmetric configurations. This equation takes the form

\begin{equation}
\label{eq:hifequation}
 \hat{\mathfrak{D}}^{2}\left(h_{I}/f\right)
-{\textstyle\frac{1}{12}}C_{IJK}
\hat{F}^{J}\cdot\hat{F}^{K}
+{\textstyle\frac{2}{\sqrt{3}}}C_{IJK}h^{J}\hat{F}^{K}\cdot G^{-}
+2g^{2}f^{-2} g_{XY} k_{I}{}^{X} k_{J}{}^{Y} h^{J}  =0\, , 
\end{equation}

\noindent
where 

\begin{equation}
G\equiv fd\omega\, .
\end{equation}

\noindent

This is the only equation that has to be solved if we have a configuration
which we know is supersymmetric and admits a gauge potential. It differs from
that of the ungauged case in the gauge-covariant derivatives and in the last
two terms. The first of these is implicitly first-order in $g$, due to
Eq.~(\ref{eq:theta-}) and the second one is manifestly second-order in $g$.

Constructing a supersymmetric configuration is, now, considerably more complex
than in the ungauged or Abelian-gauged cases and it seems not possible to give
an algorithm which automatically returns supersymmetric configurations. At any
rate, a possible recipe to construct a supersymmetric configuration of a given
$N=1,d=5$ gauged supergravity theory is the following.

\begin{enumerate}

\item The objects that have to be chosen are 

  \begin{enumerate}
  \item The 4-dimensional spatial metric $h_{\underline{m}\underline{n}}(x)$
    and an almost complex structure $\vec{\Phi}_{mn}$. The former determines
    the anti-selfdual part of its spin connection: $\xi^{-}{}_{mnp}$.
    
  \item A spatial 1-form $\omega_{\underline{m}}$. 
    
  \item The $4 n_{H}$ hyperscalar mappings $q^{X}(x)$ from the 4-dimensional
    spatial manifold to the quaternionic-K\"ahler manifold. They determine the
    (pullbacks of) the momentum map\footnote{If $n_{H}=0$ they are constant
      Fayet-Iliopoulos terms as explained in footnote~\ref{foot:esa}.}
    $\vec{P}_{I}$ and the $SU(2)$ connection
    $\vec{A}_{\underline{m}}=\partial_{\underline{m}}q^{X} \vec{\omega}_{X}$
    
  \item A spatial gauge potential $\hat{A}^{I}{}_{\underline{m}}$. It
    determines the spatial gauge field strength
    $\hat{F}^{I}{}_{\underline{m}\underline{n}}$ and, together with the
    pullback of the $SU(2)$ connection $\vec{A}_{\underline{m}}$ and the
    momentum map, it determines the spatial $SU(2)$ connection $\hat{\vec{B}}$
    whose definition we rewrite here for convenience:

    \begin{displaymath}
     \hat{\vec{B}} \equiv 
     \vec{A}+{\textstyle\frac{1}{2}} g\hat{A}^{I}\vec{P}_{I}\, .
     \end{displaymath}
     
   \item $\bar{n}=n_{V}+1$ scalar functions $h_{I}/f$. They determine, upon
     use of the constraint $C_{IJK}h^{I}h^{J} h^{L}=1$ the $n_{V}$ scalars
     $\phi^{x}$ and the metric function $f$\footnote{One can also use
       Eq.~(\ref{eq:f}) to determine $f$.}. Together with
     $\hat{A}^{I}{}_{\underline{m}}$ and $\omega_{\underline{m}}$ they give
     the full 5-dimensional gauge potential $A^{I}{}_{\mu}$

     \begin{displaymath}
     A^{I} =  -\sqrt{3} h^{I} e^{0} + \hat{A}^{I} \, . 
     \end{displaymath}

\end{enumerate}

\item These objects now have to satisfy the following equations:

  \begin{enumerate}
  \item Eq.~(\ref{eq:embedding}) that embeds the spatial $SU(2)$ connection
    $\hat{\vec{B}}$ into the spin connection of the base spatial manifold.

\begin{displaymath}
\xi^{-}{}_{mnp}  =
-\hat{\vec{B}}_{m} \cdot \vec{\Phi}_{np} 
-{\textstyle\frac{1}{4}}\partial_{m}\vec{\Phi}_{nq}\cdot\vec{\Phi}_{qp}\, ,
\end{displaymath}

  \item Eq.~(\ref{triholomorphic}) that characterizes the hyperscalar
    mappings

\begin{displaymath}
\hat{\mathfrak{D}}_{m} q^{X} 
=
\Phi^{r}{}_{m}{}^{n}\ \hat{\mathfrak{D}}_{n} q^{Y}\ J^{r}{}_{Y}{}^{X}\, .
\end{displaymath}

\item Eqs.~(\ref{remanentconstraint}) and (\ref{eq:theta-}) 

\begin{displaymath}
  \begin{array}{rcl}
h_{I}\hat{F}^{I+} & = &  
{\textstyle\frac{2}{\sqrt{3}}} (fd\omega)^{+} \, , \\
& &  \\
\hat{F}^{I-} & = & 
-2gf^{-1}C^{IJK}h_{J}\vec{P}_{K}\cdot\vec{\Phi}\, . \\
\end{array}
\end{displaymath}

\item Finally, Eq.~(\ref{eq:hifequation})

  \begin{displaymath}
 \hat{\mathfrak{D}}^{2}\left(h_{I}/f\right)
-{\textstyle\frac{1}{12}}C_{IJK}
\hat{F}^{J}\cdot\hat{F}^{K}
+{\textstyle\frac{2}{\sqrt{3}}}C_{IJK}h^{J}\hat{F}^{K}\cdot G^{-}
+2g^{2}f^{-2} g_{XY} k_{I}{}^{X} k_{J}{}^{Y} h^{J}  =0\, .     
  \end{displaymath}
  \end{enumerate}

\end{enumerate}

As we see, finding supersymmetric solutions remains a difficult problem and we
leave for future work the construction of explicit examples \cite{kn:BCMO}.

\subsection{The null case}
\label{sec-null}


\subsubsection{The equations for the bilinears}

As usual, we denote the null Killing vector by $l^{\mu}$ and choose null
coordinates $u$ and $v$ such that

\begin{equation}
l_{\mu}dx^{\mu}= f du\, ,
\hspace{1cm}
l^{\mu}\partial_{\mu} =\partial_{\underline{v}}\, ,  
\end{equation}

\noindent
where $f$ may depend on $u$ but not on $v$. The metric can be put in the form

\begin{equation}
\label{eq:nullmetric}
ds^{2}= 2fdu(dv+Hdu+\omega)
-f^{-2}\gamma_{\underline{r}\underline{s}}dx^{r}dx^{s}\, ,  
\end{equation}

\noindent
where $r,s,t=1,2,3$ and the 3-dimensional spatial metric
$\gamma_{\underline{r}\underline{s}}$ may also depend on $u$ but not on $v$.
With these coordinates the partial gauge fixing~(\ref{gaugefixing}), for
$g\neq 0$, becomes just $A^{I}_{\underline{v}}=0$. Eqs.~(\ref{lvphi}) and
(\ref{lvq}) state that the scalars are $v$-independent.

In the null case Fierz identities imply that the 2-forms bilinears $\Phi^{r}$
 are given by

\begin{equation}
\label{eq:null2-forms}
\Phi^{r} = du\wedge v^{r} \, ,
\end{equation}

\noindent
where the $v^{r}$ are 1-forms that can be used as Dreibein for the spatial
metric $\gamma_{\underline{r}\underline{s}}$.

We decompose the gauge potential as

\begin{equation}
\label{eq:potentialdecomposition}
A^{I} = A^{I}{}_{\underline{u}} du + \hat{A}^{I} \, ,
\end{equation}

\noindent
where $\hat{A}$ is a spatial one-form. Under a $u$-independent
$G$-transformation $\hat{A}^{I}$ transforms as a gauge connection whereas
$A^{I}{}_{\underline{u}}$ transforms homogeneously. We denote by
$\hat{\mathfrak{D}}$ the spatial covariant derivative made with the
three-dimensional affine and spin connections and the gauge connection
$\hat{A}^{I}$.

Eq.~(\ref{eq:covariantconstancy}) becomes

\begin{equation}
du\wedge\left[ dv^{r} - \left( 2\varepsilon^{rst} \hat{B}^{t} 
+ \sqrt{3}gf^{-1} h^{I}P_{I}^{s} v^{r}\right)\wedge v^{s} \right]
=0\, ,
\end{equation}

\noindent
where, again, $\hat{B}^{t}$ is $B^{t}$ with $A^{I}$ replaced by $\hat{A}^{I}$.
This equation relates the the tridimensional spin connection (computed for
constant $u$) to the spatial components of the pullback of the $SU(2)$:

\begin{equation}
\varpi^{rs} =  
2\varepsilon^{rst}\hat{B}^{t}
-2\sqrt{3}gf^{-1} h^{I}P_{I}^{[r} v^{s]}\, .
\label{eq:FixOmegaNull}
\end{equation}

Substituting the 2-forms we found into Eq.~(\ref{preholomorphicq}) we arrive
at

\begin{equation}
\label{eq:conditiononqx}
\hat{\mathfrak{D}}_{r} q^{X} J^{r}{}_{X}{}^{Y} = 
\sqrt{3} g f^{-1} h^{I} k_{I}{}^{Y}  \, ,
\end{equation}

\noindent
which is the condition that must be satisfied by the mappings $q^{X}$ in order
to have supersymmetry.

Let us now determine the vector field strengths: Eqs.~(\ref{df}) and
(\ref{dphi}) lead to

\begin{equation}
l^{\mu}F^{I}{}_{\mu\nu}=0\, ,  
\end{equation}

\noindent
which implies that the field strengths have the general form

\begin{equation}
F^{I} = F^{I}{}_{+r}e^{+}\wedge e^{r} 
+{\textstyle\frac{1}{2}} f^{2} F^{I}{}_{rs}e^{r}\wedge e^{s}=
    F^{I}{}_{+r}du\wedge v^{r} 
+{\textstyle\frac{1}{2}} F^{I}{}_{rs}v^{r}\wedge v^{s}\equiv 
F^{I}{}_{+r}du\wedge v^{r} 
+\hat{F}^{I}\, .
\end{equation}

\noindent
From Eq.~(\ref{dV}) we get

\begin{equation}
\label{eq:FIrsNull}
h_{I}\hat{F}^{I} =
\sqrt{3}\hat{\star}\hat{d}f^{-1}
+2gf^{-2}h^{I}\hat{\star}\hat{P}_{I}\, , 
\end{equation}

\noindent
where $\hat{P}_{I}$ is the spatial 1-form

\begin{equation}
\hat{P}_{I} = P^{r}{}_{I}v^{r}\, .  
\end{equation}

\noindent
On the other hand Eq.~(\ref{eq:Phidphi}) yields

\begin{equation}
\label{eq:FxrsNull}
h^{x}_{I}\hat{F}^{I} \; =\; 
-f^{-1}\hat{\star} \hat{\mathfrak{D}}\phi^{x}
+2gf^{-2}h^{x}_{I} \hat{\star}\hat{P}^{I}\, ,
\end{equation}

\noindent
which, together with the previous equation and the definition of $h^{x}_{I}$
give

\begin{equation}
\hat{\star} \hat{F}^{I} \; =\;  
\sqrt{3} \hat{\mathfrak{D}}(h^{I}/f)
+2gf^{-2}\hat{P}^{I}\, .
\end{equation}

From the $++r$ components of Eq.~(\ref{nablaPhi}) we get 

\begin{equation}
\label{eq:FIprNull}
h_{I}F^{I}{}_{+r} =
-{\textstyle\frac{1}{\sqrt{3}}}f^{2} (\hat{\star} F)_{r}\, ,
\end{equation}

\noindent
where

\begin{equation}
F =\hat{d}\omega\, .
\end{equation}

The components $h^{x}_{I}F^{I}{}_{+r}$ are not determined by supersymmetry and
we parametrize them by 1-forms $\psi^{I}$ satisfying $h_{I}\psi^{I}=0$.  In
conclusion, the vector field strengths must take the general form

\begin{equation}
\label{eq:vectorfieldstrengths}
F^{I} = 
({\textstyle\frac{1}{\sqrt{3}}} f^{2}h^{I}\hat{\star}F -\psi^{I})\wedge du
+\sqrt{3}\hat{\star} 
\left[
\hat{\mathfrak{D}}(h^{I}/f)
+{\textstyle\frac{2}{\sqrt{3}}}gf^{-2} \hat{P}^{I}\right]\, .
\end{equation}


\subsubsection{Solving the Killing spinor equations}
\label{sec:SolvKSENull}

It is not difficult to check that, for field configurations with metric of the
form Eq.~(\ref{eq:nullmetric}), vector field strengths of the form
Eq.~(\ref{eq:vectorfieldstrengths}) and hyperscalars satisfying
Eq.~(\ref{eq:conditiononqx}), the KSEs admit solutions which are constant
spinors satisfying the constraint

\begin{equation}
\gamma^{+}\epsilon^{i} =0\, ,
\label{eq:+projection}
\end{equation}

\noindent
and a constraint of the form 

\begin{equation}
\Pi^{r}\epsilon =  0\, ,  
\label{eq:piprojection}
\end{equation}

\noindent
for every $r$ for which $\hat{B}^{r}$ and $gfh^{I}P_{I}^{r}$ do not vanish,
where $\Pi^{r}$ is the projector 

\begin{equation}
  \label{eq:DefNullProj}
  \Pi^{r}{}_i{}^{j} \; =\; 
\textstyle{1\over 2}\left(\delta- i\gamma^{(r)}\sigma^{(r)}\right)_i{}^{j} 
  \hspace{.5cm};\hspace{.5cm} {\Pi^{r}}^{2} \ =\ \Pi^{r}
  \hspace{.5cm};\hspace{.5cm} \left[\ \Pi^{r}\ ,\ \Pi^{s}\ \right] \ =\ 0\; .
\end{equation}

\noindent 
Each of these projections breaks/preserves one half of the supersymmetries. In
the general case one must impose the three projections given in
Eq.~(\ref{eq:piprojection}). It should be noted that in this case the
projection~(\ref{eq:+projection}) is already implied by the whole system of
projections (\ref{eq:piprojection}). Thus we have that the general
supersymmetric configurations preserve $1/8$ of the supersymmetries.

As it happened in Ref.~\cite{Bellorin:2006yr} consistency with the
space-independence of the Killing spinors requires the $u$-component of $B$ to
have the form

\begin{equation}
\label{eq:NullAConsistency}
v_{[r}{}^{\underline r}\partial_{\underline{u}}v_{s]\underline{r}}
=
-2\varepsilon_{rst}B^{t}{}_{\underline{u}}\, .
\end{equation}

\subsubsection{Equations of motion}

We now want to impose the equations of motion on the supersymmetric
configurations that we have identified.  On supersymmetric configurations only
a few equations of motion are independent, since they are related by the
Killing Spinor Identities (KSIs) \cite{Kallosh:1993wx,Bellorin:2005hy} which,
as discussed in Section~\ref{sec-n1d5mg}, for these theories were computed in
Ref.~\cite{Bellorin:2006yr}.  A few KSIs were overlooked, however, in the
reference. They reduce considerably the number of independent equations to be
checked and we start by computing them.


\vspace{.5cm}
\noindent
\underline{\textit{Additional KSIs}}
\vspace{.5cm}

According to Eq.~(\ref{eq:null2-forms}) the only non-vanishing components of
the 2-forms $\Phi^{r}$ are

\begin{equation}
\Phi^{r\, s-} =\delta^{rs}\, . 
\end{equation}

We can use this result to find additional constraints in the equations of
motion from the KSIs \cite{Bellorin:2006yr}

\begin{eqnarray}
\label{eq:ksi1}
  \left[\left( \mathcal{E}_{bc}
+{\textstyle\frac{\sqrt{3}}{2}}
h_{I}\star\mathcal{B}^{I}{}_{bc}\right)
\gamma^{c} +{\textstyle\frac{\sqrt{3}}{2}}h^{I}\mathcal{E}_{I\, b}
\right]\epsilon^{i} & = & 0\, ,\\
& & \nonumber \\
\label{eq:ksi2}
\left[\mathcal{E}_{x} -h^{I}_{x} \left(\not\!\mathcal{E}_{I}
+{\textstyle\frac{1}{6}}a_{IJ}\not\!\mathcal{B}^{J}\right) \right]
\epsilon^{i} & = & 0\, .
\end{eqnarray}

Acting with $(\sigma^{r})^{j}{}_{i}\bar{\epsilon}_{j}\gamma^{a}$ on
Eq.~(\ref{eq:ksi1}), we get 

\begin{equation}
\left( \mathcal{E}_{bc}
+{\textstyle\frac{\sqrt{3}}{2}}
h_{I}\star\mathcal{B}^{I}{}_{bc}\right)
\Phi^{r\, ac}=0\, .  
\end{equation}

\noindent
Taking $a=-,r$ we get, respectively

\begin{eqnarray}
\label{eq:ksi3}
\mathcal{E}_{br}
& = & -{\textstyle\frac{\sqrt{3}}{2}}
h_{I}\star\mathcal{B}^{I}{}_{br}\, ,\\
& & \nonumber \\
\label{eq:ksi4}
\mathcal{E}_{b-}
& = & -{\textstyle\frac{\sqrt{3}}{2}}
h_{I}\star\mathcal{B}^{I}{}_{b-}\, .
\end{eqnarray}

\noindent
The second identity was already found in \cite{Bellorin:2006yr}. The symmetry
of the l.h.s.~and the antisymmetry of the r.h.s.~of both identities and the
combination of both implies

\begin{eqnarray}
\label{eq:ksi5}
\mathcal{E}_{r-} & = & h_{I}\star\mathcal{B}^{I}{}_{r-}=0 \, ,\\
& & \nonumber \\
\label{eq:ksi6}  
\mathcal{E}_{rs} & = & h_{I}\star\mathcal{B}^{I}{}_{rs}=0 \, .
\end{eqnarray}

Eqs.~(\ref{eq:ksi3})-(\ref{eq:ksi6}) leave us with only three non-vanishing
components of the Einstein equations, namely
$\mathcal{E}_{++},\mathcal{E}_{+-},\mathcal{E}_{+t}$, of which the last two
are proportional to components of the Bianchi identities. Thus, the only
independent component of the Einstein equation is $\mathcal{E}_{++}$.

Acting now with $(\sigma^{r})^{j}{}_{i}\bar{\epsilon}_{j}$ on
Eq.~(\ref{eq:ksi2}), we get 

\begin{equation}
\label{eq:ksi11}
h_{Ix}\star\mathcal{B}^{I}{}_{ab}\Phi^{r\, ab}=0\, ,\,\,\,\,
\Rightarrow \,\,
h_{Ix}\star\mathcal{B}^{I}{}_{-r}=0\, ,
\end{equation}

\noindent
which, together with Eq.~(\ref{eq:ksi5}) leads to

\begin{equation}
\label{eq:ksi12}
\star\mathcal{B}^{I}{}_{-r}=0\, .
\end{equation}

Acting  with $(\sigma^{r})^{j}{}_{i}\bar{\epsilon}_{j}\gamma^{a}$ on
Eq.~(\ref{eq:ksi2}), we get 

\begin{eqnarray}
\label{eq:ksi13}
h^{I}_{x}\mathcal{E}_{I-} & = & 0\, ,\\
& & \nonumber \\
\label{eq:ksi14}
h^{I}_{x}\mathcal{E}_{Ir} & = & 
{\textstyle\frac{1}{2}}h_{Ix}\varepsilon_{rst}\star\mathcal{B}^{I}{}_{st}\, ,
\end{eqnarray}

\noindent
which, together with $h^{I}\mathcal{E}_{I\, \mu}=0$ (proven in
Ref.~\cite{Bellorin:2006yr}) imply

\begin{equation}
\label{eq:ksi15}
\mathcal{E}_{I-}=0\, .  
\end{equation}

The only independent components of the Maxwell equations are
$h^{I}_{x}\mathcal{E}_{I+}$.

Summarizing, unbroken supersymmetry implies that the only non-automatically
vanishing components of the Einstein and Maxwell equations and Bianchi
identities are $\mathcal{E}_{++}, \mathcal{E}_{+-}, \mathcal{E}_{+r}$,
$\mathcal{B}^{I}{}_{+-}, \mathcal{B}^{I}{}_{+r}, \mathcal{B}^{I}{}_{rs}$ and
$\mathcal{E}_{I+}, \mathcal{E}_{Ir}$. The scalar equations of motion are
always automatically satisfied. If the Bianchi identities are satisfied, as
they must in this case\footnote{In the non-Abelian case that we are
  considering here the knowledge of the gauge potential is necessary in order
  to construct a supersymmetric configuration, which is our starting point,
  and the Bianchi identities are always assumed to be satisfied. Nevertheless,
  since the gauge field strength is related to other fields, the Bianchi
  identities lead to constraints on the other fields. }, only
$\mathcal{E}_{++}$ and $\mathcal{E}_{I+}$ need to be explicitly checked.


\vspace{.5cm}
\noindent
\underline{\textit{Independent equations of motion}}
\vspace{.5cm}

Let us start with the Bianchi identities.  Using the decomposition of the
potential Eq.~(\ref{eq:potentialdecomposition}) we obtain from the expression
for the gauge field strength Eq.~(\ref{eq:vectorfieldstrengths}) two
equations:

\begin{eqnarray}
\label{eq:nullbianchi1}
\hat{F}^{I} & = & 
\sqrt{3}\hat{\star} 
\left[
\hat{\mathfrak{D}}(h^{I}/f)
+{\textstyle\frac{2}{\sqrt{3}}}gf^{-2} \hat{P}^{I}\right]\, , \\ 
& & \nonumber \\ 
\label{eq:nullbianchi2}
\hat{\mathfrak{D}} A^{I}{}_{\underline{u}} 
-\partial_{\underline{u}} \hat{A}^{I}
& = &
{\textstyle\frac{1}{\sqrt{3}}} f^{2}h^{I}\hat{\star}F -\psi^{I}\, .
`\end{eqnarray}

\noindent
The Bianchi identity of the first equation leads to

\begin{equation}
\label{eq:harmonichf}
\hat{\mathfrak{D}}\hat{\star}\hat{\mathfrak{D}} (h^{I}/f) 
+{\textstyle\frac{2}{\sqrt{3}}}g 
\hat{\mathfrak{D}}(f^{-2} \hat{\star}\hat{P}^{I})=0\, .
\end{equation}

\noindent
The constraint $h_{I} \psi^{I} = 0$ and the second equation imply

\begin{equation}
\label{eq:omega}
{\textstyle\frac{1}{\sqrt{3}}} f^{2} \hat{\star}F 
-h_{I} \hat{\mathfrak{D}} A^{I}{}_{\underline{u}} 
+ h_{I} \partial_{\underline{u}}\hat{A}^{I} 
=0\, ,
\end{equation}

\noindent
which can be taken as the equation defining $\omega$. Having $\omega$ and the
potentials Eq.~(\ref{eq:nullbianchi2}) determines $\psi^{I}$:

\begin{equation}
\label{eq:psi}
\psi^{I} = {\textstyle\frac{1}{\sqrt{3}}} f^{2}h^{I}\hat{\star}F 
-\hat{\mathfrak{D}} A^{I}{}_{\underline{u}} 
+ \partial_{\underline{u}} \hat{A}^{I}\, .
\end{equation}

Apart from these equations we have to impose the Maxwell equations, which, in
differential form language take the form

\begin{equation}
4\star\mathcal{E}_{I} =
 -\mathfrak{D}\star\left(a_{IJ}F^{J}\right)
+{\textstyle\frac{1}{\sqrt{3}}}C_{IJK}F^{J}\wedge F^{K}
+g \star \left(
k_{I\,x} \mathfrak{D}\phi^{x} + k_{I\,X} \mathfrak{D} q^{X}
\right)\, .
\end{equation}

Substituting the gauge field strength and operating we get 

\begin{equation}
\label{eq:Maxwellenbruto}
  \begin{array}{rcl}
4\star\mathcal{E}_{I} & = &  du\wedge 
\left\{
g\left[\sqrt{3}f_{IJ}{}^{K}\hat{F}^{J}h_{K}f 
-2\hat{\mathfrak{D}}\hat{P}_{I} 
-\hat{\star}(k_{I\,x} \hat{\mathfrak{D}}\phi^{x} 
+k_{I\,X} \hat{\mathfrak{D}} q^{X})\right]
\wedge (dv+\omega) 
\right. \\
& & \\
& & 
-\sqrt{3}
\left[ 
\hat{\mathfrak{D}}(h_{I}f)
-{\textstyle\frac{2}{\sqrt{3}}}g \hat{P}_{I} 
\right]\wedge F
+{\textstyle\frac{1}{\sqrt{3}}}\hat{\mathfrak{D}}(fh_{I})\wedge F
-\hat{\mathfrak{D}}(f^{-1}\hat{\star}\psi_{I})
\\
& & \\
& & 
\left.
-g f^{-3}\hat{\star}\left(
k_{I\,x} \mathfrak{D}_{\underline{u}}\phi^{x} 
+k_{I\,X} \mathfrak{D}_{\underline{u}} q^{X}
\right) 
-{\textstyle\frac{2}{\sqrt{3}}}  C_{IJK} 
({\textstyle\frac{1}{\sqrt{3}}} f^{2}h^{J}\hat{\star}F -\psi^{J})
\wedge \hat{F}^{K}
\right\}\, .
\end{array}
\end{equation}

The first line contributes to $\mathcal{E}_{I\, r}$ and it can be checked
(thorugh a long and painful calculation) that it vanishes automatically for
supersymmetric configurations, as it should according to the KSIs, while the
other two lines contribute to $\mathcal{E}_{I\, +}$.

The Maxwell equations, then, simplify and take the form

\begin{equation}
  \begin{array}{rcl}
4\star\mathcal{E}_{I} & = &  du\wedge 
\left\{
\left[ 
-\sqrt{3}f\hat{\mathfrak{D}}(h_{I})
+2g \hat{P}_{I} 
-{\textstyle\frac{4}{3}} gC_{IJK}h^{J}\hat{P}^{K}
\right] 
\wedge F
-\hat{\mathfrak{D}}(\hat{\star}\psi_{I}/f)
\right.
\\
& & \\
& & 
\left.
+{\textstyle\frac{2}{\sqrt{3}}}  C_{IJK} \psi^{J}\wedge \hat{F}^{K}
-g f^{-3}\hat{\star}\left(
k_{I\,x} \mathfrak{D}_{\underline{u}}\phi^{x} 
+k_{I\,X} \mathfrak{D}_{\underline{u}} q^{X}
\right) 
\right\}\, .
\end{array}
\end{equation}

\noindent
As implied by the KSIs only the $\mathcal{E}_{I+}$ component is not
automatically satisfied and has to be explicitly imposed in order to get
classical solutions. It can be also be checked that $h^{I}\mathcal{E}_{I+}=0$
(as it is implied by the KSIs) up to terms that are proportional to
$d^{2}\omega$.

The same fact can be described in a slightly different way: the integrability
condition of the $\omega$ equation ($d^{2}\omega=0$) is satisfied if
supersymmetry is unbroken and the KSI $h^{I}\mathcal{E}_{I+}=0$ is satisfied.
In general, as first pointed out in Refs.~\cite{Denef:2000nb,Bates:2003vx}
there will be singular points at which this will not happen. These points give
rise to physical singularities in the metric and, therefore, they should not
be allowed in meaningful solutions. This requirement translates into
constraints on charges and asymptotic values of the moduli.  It can be argued
that this requirement is equivalent to the requirement of having supersymmetry
unbroken everywhere (and the KSIs satisfied everywhere)
\cite{Bellorin:2006xr,Ortin:2006xg}.

In order to write the equations of motion in a simple form it is convenient to
define some new variables:

\begin{eqnarray}
h^{I}/f & \equiv & K^{I}\, ,\,\,\,\,\,
f^{-3} = C_{IJK} K^{I} K^{J} K^{K}\, , \\
& & \nonumber \\
L_{I} & \equiv & C_{IJK} K^{J} A^{K}{}_{\underline{u}}\, ,\\
& & \nonumber \\
N & \equiv & H+{\textstyle\frac{1}{2}} L_{I} A^{I}{}_{\underline{u}}\, .
\end{eqnarray}

Observe that $\frac{1}{\sqrt{3}} \hat{A}^{I}$ and $-A^{I}{}_{\underline{u}}$
coincide, respectively, with what was called $\alpha^{I}$ and $M^{I}$ in the
ungauged case, in Ref.~\cite{Bellorin:2006yr}.

Using these variables and Eq.~(\ref{eq:psi}), the Maxwell equation can be put
into the form

\begin{equation}
\label{eq:maxwellfinal}
  \begin{array}{rcl}
4\star\mathcal{E}_{I} & = &  -2du\wedge 
\left\{
\hat{\mathfrak{D}}\hat{\star}\hat{\mathfrak{D}}L_{I}
-g \hat{P}_{I} \wedge F
+{\textstyle\frac{2}{\sqrt{3}}}g 
C_{IJK}
\hat{\mathfrak{D}}
\hat{\star}(f^{-2}A^{J}{}_{\underline{u}}\hat{P}^{K})
\right.
\\
& & \\
& & 
-gC_{IJK}
\left[\hat{\mathfrak{D}}\hat{\star}(K^{J}\partial_{\underline{u}}\hat{A}^{K})
+
(\hat{\mathfrak{D}}K^{J}
+{\textstyle\frac{2}{\sqrt{3}}}g \hat{\star}\hat{P}^{J})
\wedge 
\hat{\star}\partial_{\underline{u}}\hat{A}^{K}
\right]
\\
& & \\
& & 
\left.
+{\textstyle\frac{1}{2}}g f^{-3}\hat{\star}\left(
k_{I\,x} \mathfrak{D}_{\underline{u}}\phi^{x} 
+k_{I\,X} \mathfrak{D}_{\underline{u}} q^{X}
\right) 
\right\}\, .
\end{array}
\end{equation}

This equation is gauge-invariant, in particular, under $u$-dependent $G$-gauge
transformations that act on $\hat{A}^{I}$, $A^{I}_{\underline{u}}$, $L_{I}$
and the bosonic scalars. This fact can be used to partially fix the $G$
gauge, as done in Ref.~\cite{Bellorin:2006yr}, leaving a much simpler equation
which is still covariant under $u$-independent $G$ gauge transformations. 

The 1-form $\omega$ is determined by Eq.~(\ref{eq:omega}) only up to total
derivatives which correspond to shifts in the coordinate $v$. This
transformation must be accompanied with a shift in $H$ (or $N$). We can use
this freedom to impose a condition on (basically, the $u$-dependence of)
$\omega$:

\begin{equation}
\begin{array}{l}
\hspace*{-10mm}
\nabla_{r}(\dot\omega)_{r} + 3(\dot\omega)_{r}\partial_{r} \log f  = 
\\ \\
- {\textstyle\frac{1}{2}}f^{-3}(\ddot\gamma)_{rr}
- {\textstyle\frac{1}{4}}f^{-3}(\dot\gamma)^2\nonumber
+ {\textstyle\frac{3}{2}} f^{-4}\dot f(\dot\gamma)_{rr}
+ 3f^{-3}[\partial_{\underline{u}}^2 \log f -2(\partial_{\underline{u}}\log f)^2]
\\ \\
-{\textstyle\frac{1}{2}}f^{-3}\left[ 
 g_{xy} (\dot \phi^{x} \dot \phi^{y}  
+2g\dot{q}^{x} A^{I}{}_{\underline{u}} k_{I}{}^{y} )
+g_{XY} (\dot q^{X} \dot q^{Y}  
+ 2g\dot{q}^{X} A^{I}{}_{\underline{u}} k_{I}{}^{Y} ) \right]
\\   \\ 
+C_{IJK} K^{I} \left[
 (\partial_{\underline{u}}\hat{A}^{J})_{r} 
(\partial_{\underline{u}}\hat{A}^{K})_{r}
-2 \hat{\mathfrak{D}}_{r} A^{J}{}_{\underline{u}} 
(\partial_{\underline{u}}\hat{A}^{K})_{r} \right]
\, .  
\end{array}
\label{eq:fixomega} 
\end{equation}

After performing these steps, the $\mathcal{E}_{++}$ component of the Einstein equations becomes

\begin{equation}
\label{eq:einstein++}
-f^{-1}\mathcal{E}_{++} =
\nabla^{2} N
+{\textstyle\frac{1}{\sqrt{3}}} g\hat{\mathfrak{D}}_{r} (f^{-2}C_{IJK}P^{I}_{r}
 A^{J}_{\underline{u}} A^{K}{}_{\underline{u}})
+{\textstyle\frac{1}{2}}gf^{-3}A^{I}{}_{\underline{u}} A^{J}{}_{\underline{u}} 
 (g_{xy}k_{I}{}^{x} k_{J}{}^{y} + g_{XY} k_{I}{}^{X} k_{J}{}^{Y} )\,.
\end{equation}

Let us summarize the results of this section by giving the ``recipe'' to build
supersymmetric solutions in the null class.

\begin{enumerate}
\item The objects that have to be chosen are

  \begin{enumerate}

  \item A spatial 3-dimensional metric $\gamma_{\underline{r}\underline{s}}$
    and Dreibein basis $v^{r}$ both of which may depend on the null coordinate
    $u$. This determines the 3-dimensional spin connection $\varpi^{rs}$.
    
  \item The $4 n_{H}$ hyperscalar $u$-dependent mappings $q^{X}(x,u)$ from the
    3-dimensional spatial manifold to the quaternionic-K\"ahler manifold. They
    determine the (pullbacks of) the momentum map $\vec{P}_{I}$ and the
    $SU(2)$ connection $\vec{A}_{\underline{r}}=\partial_{\underline{r}}q^{X}
    \vec{\omega}_{X}$ and
    $\vec{A}_{\underline{u}}=\partial_{\underline{u}}q^{X} \vec{\omega}_{X}$
  \item A gauge connection 1-form $A^{I}$ with vanishing $v$ component. This
    determines its spatial and null parts $\hat{A}^{I}$ and
    $A^{I}{}_{\underline{u}}$.
  \item $2\bar{n}+1$ functions $K^{I},L_{I},N$. They determine the functions
    $f,K_{I}$ and $H$, and, together with $\omega,\hat{A}^{I}$ and
    $A^{I}{}_{\underline{u}}$, the 1-forms $\psi$ via Eq.~(\ref{eq:psi}) and
    the spatial 1-form $\omega$ via Eq.~(\ref{eq:omega}) which can be written
    in the form

\begin{equation}
\hat{\star} F = \sqrt{3} (K^{I}\hat{\mathfrak{D}} L_{I} 
-L_{I} \hat{\mathfrak{D}} K^{I})
-\sqrt{3} K_{I}\partial_{\underline{u}}\hat{A}^{I}\,.
\end{equation}

  \end{enumerate}

\item These objects must satisfy the following equations:

  \begin{enumerate}
  \item Eq.~(\ref{eq:conditiononqx}) that characterizes the quaternionic
    mappings $q^{X}$ and relates them to the spatial components of the gauge
    connection $\hat{A}^{I}$ and the functions $K^{I}$:

\begin{equation}
\hat{\mathfrak{D}}_{r} q^{X} J^{r}{}_{X}{}^{Y} = 
\sqrt{3} g K^{I} k_{I}{}^{Y}  \, .
\end{equation}

\item Eq.~(\ref{eq:FixOmegaNull}) which relates the spatial components of the
  pullback of the $SU(2)$ connection with the 3-dimensional spin connection,
  the spatial components of the gauge connection $\hat{A}^{I}$ and the
  functions $K^{I}$:

\begin{equation}
\varpi^{rs} =  
2\varepsilon^{rst}\hat{B}^{t}
-2\sqrt{3}g K^{I}P_{I}^{[r} v^{s]}\, .
\end{equation}

\item Eq.~(\ref{eq:NullAConsistency}) which relates the null component of the
  pullback of the $SU(2)$ connection with the Dreibeins and the null
  components of the gauge connection $A^{I}{}_{\underline{u}}$:

\begin{equation}
v_{[r}{}^{\underline r}\partial_{\underline{u}}v_{s]\underline{r}}
=
-2\varepsilon_{rst}B^{t}{}_{\underline{u}}\, .
\end{equation}

\item Eq.~(\ref{eq:harmonichf}), which follows from the Bianchi identity and
  can be put in the form

\begin{equation}
\hat{\mathfrak{D}}\hat{\star}\hat{\mathfrak{D}} K^{I} 
+ {\textstyle\frac{2}{\sqrt{3}}}g 
\hat{\mathfrak{D}}(\hat{\star}f^{-2} \hat{P}^{I})
=0\,.
\label{eq:harmonicK}
\end{equation}

\item Eq.~(\ref{eq:maxwellfinal}), the only independent Maxwell equation

\begin{equation}
  \begin{array}{rcl}
\hat{\mathfrak{D}}\hat{\star}\hat{\mathfrak{D}}L_{I}
-g \hat{P}_{I} \wedge F
+{\textstyle\frac{2}{\sqrt{3}}}g 
C_{IJK}
\hat{\mathfrak{D}}
\hat{\star}(f^{-2}A^{J}{}_{\underline{u}}\hat{P}^{K})
& & \\
& & \\
-gC_{IJK}
\left[\hat{\mathfrak{D}}\hat{\star}(K^{J}\partial_{\underline{u}}\hat{A}^{K})
+
(\hat{\mathfrak{D}}K^{J}
+{\textstyle\frac{2}{\sqrt{3}}}g \hat{\star}\hat{P}^{J})
\wedge 
\hat{\star}\partial_{\underline{u}}\hat{A}^{K}
\right]
& & \\
& & \\
+{\textstyle\frac{1}{2}}g f^{-3}\hat{\star}\left(
k_{I\,x} \mathfrak{D}_{\underline{u}}\phi^{x} 
+k_{I\,X} \mathfrak{D}_{\underline{u}} q^{X}
\right)
& = & 0\, .\\
\end{array}
\end{equation}

\item Eq.~(\ref{eq:einstein++}), the only independent component of the Einstein
  equations:

\begin{equation}
\hat{d}\hat{\star}\hat{d}\, N
-{\textstyle\frac{1}{\sqrt{3}}} g\hat{\mathfrak{D}}\hat{\star} 
(f^{-2}C_{IJK}\hat{P}^{I} A^{J}_{\underline{u}} A^{K}{}_{\underline{u}})
+{\textstyle\frac{1}{2}}g\hat{\star}
f^{-3}A^{I}{}_{\underline{u}} A^{J}{}_{\underline{u}} 
 (g_{xy}k_{I}{}^{x} k_{J}{}^{y} + g_{XY} k_{I}{}^{X} k_{J}{}^{Y} )=0\,.
\end{equation}

  \end{enumerate}
\end{enumerate}







\section{Conclusions}
\label{sec-conclusions}

We have succeeded in finding a set of conditions which are necessary and
sufficient for a configuration of gauged $N=1,d=5$ supergravity coupled to
vector multiplets and hypermultiplets to be, first, supersymmetric and,
second, a supersymmetric classical solution. As announced in the Introduction,
the equations that we have obtained are highly non-linear and coupled, which
does not seem to allow a systematic construction of non-trivial supersymmetric
solutions.  We leave the construction and study of examples for a future
publication \cite{kn:BCMO}.

On the other hand, there exists an alternative supermultiplet for minimal
supergravity in $d=5$ \cite{Nishino:2000cz,Nishino:2001ji,Fujita:2001kv}. it
would be interesting to study the relations between the supersymmetric
configurations we have found and those of the alternative formulation.


\section*{Acknowledgments}

T.O.~would like to thank M.M.~Fern\'andez for her continuous support.  This
work has been supported in part by the Spanish Ministry of Science and
Education grant FPA2006-00783, the Comunidad de Madrid grant HEPHACOS
P-ESP-00346 and by the EU Research Training Network \textit{Constituents,
  Fundamental Forces and Symmetries of the Universe} MRTN-CT-2004-005104.

\appendix

\section{The gauging of isometries of the scalar manifolds}
\label{sec-gauging}

In this appendix we are going to review briefly the gauging of the isometries
of the scalar manifolds of $N=1,d=5$ supergravity in order to clarify some
definitions and conventions. This material is covered in a slightly different
for in Refs.~\cite{Bergshoeff:2002qk} and \cite{Bergshoeff:2004kh}.


\subsection{Killing vectors and gauge transformations}

The complete scalar manifold (or target space) of the scalar fields of
$N=1,d=5$ supergravity is the product of a real special manifold and a
quaternionic K\"ahler manifold parametrized, respectively, by the scalars of
the vector supermultiplets ($\phi^{x}$) and by the scalars of the
hypermultiplets ($q^{X}$). The metrics of these two manifolds are denoted by
$g_{xy}(\phi)$ and $g_{XY}(q)$.

We can describe the most general $N=1,d=5$ gauged supergravity theory by
focusing on the gauging of the isometries of the scalar manifolds. In the end
we will see that there are gaugings (necessarily Abelian) unrelated to
isometries that fit in the general description.

The isometries to be gauged are generated by Killing vectors of the real
special manifold $k_{I}{}^{x}(\phi) \partial_{x}$ and the quaternionic
K\"ahler manifold $k_{I}{}^{X}(q)\partial_{X}$, a pair for each vector
$A^{I}{}_{\mu}$ of the theory, although some (or all) can be identically zero.

The isometries generated by the Killing vectors $k_{I}{}^{X}$ act on the
quaternions according to

\begin{equation}
\delta_{\Lambda} q^{X} = -g\Lambda^{I} k_{I}{}^{X}\, .
\end{equation}

In the gauged theory the $\Lambda^{I}$s are the local parameters of vector
gauge transformations

\begin{equation}
\delta_{\Lambda} A^{I}{}_{\mu} = \partial_{\mu}\Lambda^{I} 
+gf_{JK}{}^{I} A^{J}{}_{\mu} \Lambda^{K}\, ,
\end{equation}

\noindent
where $f_{JK}{}^{I}$ are the structure constants of the gauge group $G$ and
are given by the Lie brackets of the $k_{I}{}^{X}$s

\begin{equation}
[k_{I},k_{J}] = -f_{IJ}{}^{K} k_{K}\, .
\end{equation}

\noindent
This implies that the functions $h^{I}$ of the real special manifold transform
in the adjoint representation of $G$:

\begin{equation}
\label{eq:deltahi}
\delta_{\Lambda} h^{I} = -gf_{JK}{}^{I}\Lambda^{J} h^{K}\, .
\end{equation}

\noindent
In turn, this implies for the scalars themselves 

\begin{equation}
\label{eq:deltaphix}
\delta_{\Lambda}\phi^{x} = -g\Lambda^{I} k_{I}{}^{x}\, ,  
\end{equation}

\noindent
where

\begin{equation}
\label{eq:kix}
k_{I}{}^{x} = -\sqrt{3} f_{IJ}{}^{K} h^{J} h_{K}^{x}\, . 
\end{equation}

These objects must be Killing vectors of $g_{xy}(\phi)$ if the $\Lambda^{I}$
transformations are also symmetries of the corresponding $\sigma$ model.
Writing $g_{xy}\partial \phi^{x} \partial \phi^{y}= -2 C_{IJKL} h^{I} \partial
h^{J}\partial h^{K}$ it is easy to see that necessary and sufficient condition
is

\begin{equation}
\label{eq:invariantmetric}
f_{I(J}{}^{K}C_{MN)K}=0\, ,  
\end{equation}

\noindent
i.e.~that $C_{IJK}$ is an invariant tensor.

Furthermore, the Killing vectors $k_{I}{}^{x}(\phi)$ satisfy the same Lie
algebra as the $k_{I}{}^{X}(q)$s and, using Eq.~(\ref{eq:invariantmetric}),
which implies 

\begin{equation}
f_{IJ}{}^{K}h^{J}h_{K}=0\, , 
\end{equation}

\noindent
they can also be written in the equivalent form

\begin{equation}
k_{I}{}^{x} 
= -\sqrt{3} f_{IJ}{}^{K} h^{Jx} h_{K}\, .
\end{equation}

The $G$-covariant derivatives on the scalars are

\begin{eqnarray}
\mathfrak{D}_{\mu} \phi^{x} & = & 
\partial_{\mu} \phi^{x}+gA^{I}{}_{\mu} k_{I}{}^{x}\, ,
\hspace*{10mm}
\mathfrak{D}_{\mu} h^{I}=
\partial_{\mu} h^{I} +gf_{JK}{}^{I} A^{J}{}_{\mu} h^{K} \, ,\\
& & \nonumber \\ 
\mathfrak{D}_{\mu} q^{X} & = & 
\partial_{\mu} q^{X} +g A^{I}{}_{\mu} k_{I}{}^{X}\, ,
\end{eqnarray}

\noindent
and they transform covariantly as

\begin{equation}
\delta_{\Lambda} \mathfrak{D}_{\mu} \varphi^{\tilde x} = 
-g\Lambda^{I}\partial_{\tilde y} k_{I}{}^{\tilde{x}} 
\mathfrak{D}_{\mu} \varphi^{\tilde{y}}\, ,
\hspace{1cm}
\delta_{\Lambda} \mathfrak{D}_{\mu} h^{I} = 
-g f_{JK}{}^{I}\Lambda^{J}  \mathfrak{D}_{\mu} h^{K}\, ,
\end{equation}

\noindent
where we have unified the notation on the scalars, $\varphi^{\tilde{x}} =
(\phi^{x},q^{X})$, $k_{I}{}^{\tilde{x}} = (k_{I}{}^{x},k_{I}{}^{X})$. 

For the sake of completeness we also quote the formulae

\begin{equation}
\mathfrak{D}_{\mu} h_{I}=
\partial_{\mu} h_{I} +gf_{IJ}{}^{K} A^{J}{}_{\mu} h_{K} \, ,
\hspace{1cm}
\mathfrak{D}_{\mu}C_{IJK}=0\, .  
\end{equation}

The second derivatives are defined by 

\begin{equation}
\mathfrak{D}_{\mu} \mathfrak{D}_{\nu} \varphi^{\tilde{x}} \equiv 
 \nabla_{\mu} \mathfrak{D}_{\nu} \varphi^{\tilde{x}}
+\Gamma_{\tilde{y}\tilde{z}}{}^{\tilde{x}}
\mathfrak{D}_{\mu}\varphi^{\tilde{y}}\mathfrak{D}_{\mu}\varphi^{\tilde{z}}
+ gA^{I}{}_{\mu} \partial_{\tilde{y}} k_{I}{}^{\tilde{x}}
\mathfrak{D}_{\nu} \varphi^{\tilde{y}} \, ,  
\end{equation}

\noindent
where $\Gamma_{\tilde{y}\tilde{z}}{}^{\tilde{x}}$ are the target space
Christoffel symbols. Their transformations and commutator are given by

\begin{eqnarray}
\delta_{\Lambda} \mathfrak{D}_{\mu} \mathfrak{D}_{\nu} \varphi^{\tilde{x}} 
& = & -g\Lambda^{I}\partial_{\tilde{y}}k_{I}{}^{\tilde{x}} 
\mathfrak{D}_{\mu} \mathfrak{D}_{\nu} \varphi^{\tilde{y}}\, , \\
& &  \nonumber \\
\left[ \mathfrak{D}_{\mu},\mathfrak{D}_{\nu} \right] \varphi^{\tilde{x}} & = &  
gF^{I}{}_{\mu\nu} k_{I}{}^{\tilde{x}}\, ,
\end{eqnarray}

\noindent
where $F^{I}{}_{\mu\nu}$ is the gauge field strength

\begin{equation}
F^{I}{}_{\mu\nu} = 2\partial_{[\mu}A^{I}{}_{\nu]}
+gf_{JK}{}^{I}A^{J}{}_{\mu}A^{K}{}_{\nu}\, .  
\end{equation}

All these definitions are enough to construct a gauge-invariant action for the
scalars, since this essentially depends on the target space metric. However,
they are not enough to gauge the full supergravity theory, which depends on
other structures as well. In particular, it depends on the complex structures
of the hyperscalar manifold and we have to study under which conditions they
are preserved by the gauging.


\subsection{The covariant Lie derivative and the momentum map}
\label{app-momentummap}

This appendix concerns only to the hyperscalar sector of the target manifold.
The quaternionic K\"ahler geometry of this manifold is defined not only by the
metric $g_{XY}$ but by the quaternionic structure $\vec{J}_{X}{}^{Y}$, which
should also be preserved by the symmetries to be gauged. Therefore, one must
require the vanishing of the Lie derivative of the  quaternionic structure
with respect to the Killing vectors $k_{I}{}^{X}$. One has to use an
$SU(2)$-\textit{covariant} Lie derivative for consistency or, as it is usually
done in the literature, impose the vanishing of the standard Lie derivative up
to gauge transformations. Here we will use an $SU(2)$-covariant Lie
derivative whose construction we describe first.

Let $\vec{\psi}$ by an $SU(2)$ vector and, simultaneously an arbitrary tensor
on the hyperscalar variety, and $\vec{\omega}$ the $SU(2)$ connection. Under
infinitesimal $SU(2)$ gauge transformations 

\begin{equation}
\delta_{\lambda}\vec{\psi} = -2\vec{\lambda}(q)\times\vec{\psi}\, ,
\hspace{10mm}
\delta_{\lambda}\vec{\omega} = -2\vec{\lambda}(q)\times\vec{\omega}
+d\vec{\lambda}(q)\, .
\end{equation}

The standard Lie derivative of $\vec{\psi}$ along the vector $k_{I}{}^{X}$
(denoted by $\mathcal{L}_{I}\vec{\psi}$) transforms under $SU(2)$ as

\begin{equation}
\delta_{\lambda}\mathcal{L}_{I}\vec{\psi}
=
-2\vec{\lambda}\times\mathcal{L}_{I}\vec{\psi}
-2\partial_{I}\vec{\lambda}\times\vec{\psi}\,,
\end{equation}

\noindent
where $\partial_{I}\equiv k_{I}{}^{X}\partial_{X}$. We now want to find
another definition of Lie derivative that transforms without derivatives of
the transformation parameter. Introducing for each Killing
vector\footnote{Only covariant Lie derivatives with respect to Killing vectors
  can be properly defined.}  $k_{I}{}^{X}$ a $\vec{\eta}_{I}$ transforming as

\begin{equation}
\label{eq:btransformation}
\delta_{\lambda}\vec{\eta}_{I} =
-2\vec{\lambda}\times\vec{\eta}_{I}+\partial_{I}\vec{\lambda}\, ,
\end{equation}

\noindent
we define the $SU(2)$-covariant Lie derivative on $SU(2)$ vectors

\begin{equation}
\mathbb{L}_{I} \vec{\psi} 
\equiv 
\mathcal{L}_{I}\vec{\psi} + 2\vec{\eta}_{I}\times\vec{\psi}\, .
\label{eq:covariantlie}
\end{equation}

For this to be a good definition  $\mathbb{L}_{I}$ must satisfy the standard
properties of a Lie derivative.

$\mathbb{L}_{I}$ is clearly a linear operator and it satisfies the Leibnitz
rule for products of $SU(2)$ vectors such as $\vec{\psi}\cdot\vec{\phi}$ and
$\vec{\psi}\times\vec{\phi}$. The Lie derivative must also satisfy

\begin{equation}
[\mathbb{L}_{I},\mathbb{L}_{J}] = \mathbb{L}_{[k_{I},k_{J}]}\, ,
\label{eq:jacobi}
\end{equation}

\noindent
which implies the Jacobi identity.  This requires the ``curvature'' of the
``connection'' $\vec{\eta}_{I}$ to be

\begin{equation}
\label{eq:etacurvature}
\partial_{I}\vec{\eta}_{J}-\partial_{J}\vec{\eta}_{I}
+2\vec{\eta}_{I}\times\vec{\eta}_{J}
=
-f_{IJ}{}^{K}\vec{\eta}_{K}\, .
\end{equation}

It should be clear that $\vec{\eta}_{I}$ must be related with the $SU(2)$
connection $\vec{\omega}$, but it is not just $k_{I}{}^{X}\vec{\omega}_{X}$,
which has the right transformation property Eq.~(\ref{eq:btransformation}) but
does not satisfy curvature property Eq.~(\ref{eq:etacurvature}). Thus, we
introduce yet another $SU(2)$ vector\footnote{We put the $-1/2$ factor to
  agree with the conventions of Ref.~\cite{Bergshoeff:2004kh}}

\begin{equation}
\vec{\eta}_{I}
= k_{I}{}^{X}\vec{\omega}_{X}-{\textstyle\frac{1}{2}}\vec{P}_{I}\, ,
\label{eq:bomegaP}
\end{equation}

\noindent
which must satisfy 

\begin{equation}
\mathfrak{D}_{I}\vec{P}_{J} - \mathfrak{D}_{J}\vec{P}_{I}
-\vec{P}_{I}\times\vec{P}_{J}
+{\textstyle\frac{1}{2}}k_{I}{}^{X}\vec{J}_{XY}k_{J}{}^{Y}
=
f_{IJ}{}^{K}\vec{P}_{K}\, ,
\label{eq:prePtimesP}
\end{equation}

\noindent
in order to meet Eq.~(\ref{eq:etacurvature}). Here we have used the fact that
in quaternionic K\"ahler manifolds the curvature of the $SU(2)$ connection is
non-vanishing and proportional to the K\"ahler two-forms. We are going to show
that $\vec{P}_{I}$ satisfies the equation that defines it as a
\textit{momentum map}.

Now, assuming that a $\vec{P}_{I}$ satisfying Eq.~(\ref{eq:prePtimesP}) has
been found, we can write the conditions that the vector $k_{I}{}^{X}$ must
satisfy to be the generator of a symmetry of the hyperscalar manifold in the
form

\begin{eqnarray}
\mathbb{L}_{I}g_{XY} &=& 0\,,
\\ \nonumber \\
\mathbb{L}_{I}\vec{J}_{XY} &=& 0\,.
\label{eq:triholomorphic}
\end{eqnarray}

\noindent
The first equation is just the Killing equation since
$\mathbb{L}_{I}g_{XY}=\mathcal{L}_{I} g_{XY}$. Given the metric and quaternionic
structure, the second condition (\textit{tri-holomorphicity} of the Killing
vectors) can be seen as a condition for $\vec{P}_{I}$ just as the Killing
equation can be seen as a condition for $k_{I}$ once the metric $g_{XY}$ is
given: it can be written in the form

\begin{equation}
-\vec{J}_{X}{}^{Y}\times\vec{P}_{I}
=
\nabla_{X}k_{I}{}^{Z} \vec{J}_{Z}{}^{Y} 
-\vec{J}_{X}{}^{Z}\nabla_{Z}k_{I}{}^{Y}\, ,
\label{eq:commutatorJdk}
\end{equation}

\noindent
which says that $\vec{P}_{I}$ measures the commutator between the quaternionic
structure and the covariant derivative of the Killing vectors. By contracting
this equation with $\vec{J}_{Y}{}^{X}$ we obtain an expression for
$\vec{P}_{I}$ itself, valid for $n_{H}\neq 0$\footnote{\label{foot:esa}In
  absence of hypermultiplets ($n_{H}=0$) the momentum map $\vec{P}_{I}$ can
  still be defined in two cases in which they are equivalent to a set of
  constant Fayet-Iliopoulos terms. In the first case the gauge group contains
  an $SU(2)$ factor and
\begin{equation}
\vec{P}_{I} = \vec{e}_{I}\,\xi\, ,
\end{equation}
where $\xi$ is an arbitrary constant and the $\vec{e}_{I}$ are constants that
are nonzero for $I$ in the range of the $SU(2)$ factor and satisfy
\begin{equation}
\vec{e}_{I}\times \vec{e}_{J} = f_{IJ}{}^{K}\vec{e}_{K}\, .  
\end{equation}
In the second case the gauge group contains a $U(1)$ factor  and 
\begin{equation}
\vec{P}_{I} = \vec{e}\,\xi_{I}\, ,
\end{equation}
where $\vec{e}$ is an arbitrary $SU(2)$ vector and the $\xi_{I}$s are
arbitrary constants that are nonzero for $I$ corresponding to the $U(1)$
factor.
} 

\begin{equation}
2n_{H}\vec{P}_{I} = \vec{J}_{X}{}^{Y}\nabla_{Y}k_{I}{}^{X}\, .
\label{eq:momentummap}
\end{equation}

For this solution to be consistent, it has to satisfy
Eq.~(\ref{eq:prePtimesP}).  To see it we first take the derivative of the
above solution Eq.~(\ref{eq:momentummap}) using the following identity for
Killing vectors,

\begin{equation}
\nabla_{X}\nabla_{Y} k^{Z} 
= 
R_{XWY}{}^{Z}k^{W}\, ,
\label{eq:killingidentity}
\end{equation}

\noindent
and the canonical decomposition of the curvature between its $SU(2)$ and
$Sp(n_{H})$ parts,

\begin{equation}
R_{XWY}{}^{Z} 
= 
-\vec{J}_{Y}{}^{Z}\cdot\vec{\mathcal{R}}_{XW} 
+ f_{Y}{}^{iB} f_{iA}{}^{Z} \mathcal{R}_{XW\,B}{}^{A}\, . 
\label{eq:curvaturedecomposition}
\end{equation}

\noindent
Only the $SU(2)$ part of the curvature contributes to the derivative of
$\vec{P}_{I}$:

\begin{equation}
\label{eq:defmomentummap}
\mathfrak{D}_{X}\vec{P}_{I} 
=
2\vec{\mathcal{R}}_{XY}k_{I}{}^{Y}
= 
-{\textstyle\frac{1}{2}}\vec{J}_{XY}k_{I}{}^{Y}\, .
\end{equation}

This equation can alternatively be taken as the definition of $\vec{P}_{I}$.
It defines a momentum map and it is crucial for coupling hypermultiplets to
supergravity.  Observe that the
integrability condition of Eq.~(\ref{eq:defmomentummap}) is precisely 
Eq.~(\ref{eq:commutatorJdk}).  

We can now substitute Eq.~(\ref{eq:defmomentummap}) in
Eq.~(\ref{eq:prePtimesP}), obtaining

\begin{equation}
\vec{P}_{I}\times\vec{P}_{J}
+{\textstyle\frac{1}{2}}k_{I}{}^{X}\vec{J}_{XY}k_{J}{}^{Y}
=
f_{IJ}{}^{K}\vec{P}_{K}\,.
\label{eq:PtimesP}
\end{equation}

\noindent
On the other hand, contracting Eq.~(\ref{eq:commutatorJdk}) with
$\nabla_{Y}k_{J}{}^{X}$ we get

\begin{equation}
n_{H}\vec{P}_{I}\times\vec{P}_{J}
=
-\vec{J}_{X}{}^{Y} \nabla_{Y}k_{[I|}{}^{Z} \nabla_{Z}k_{|J]}{}^{X}\,,
\end{equation}

\noindent
integrating by parts the right hand side of this expression, using the algebra
of the Killing vectors, identity~(\ref{eq:killingidentity}), the Bianchi
identity of the curvature and the curvature
decomposition~(\ref{eq:curvaturedecomposition}) one recovers
Eq.~(\ref{eq:PtimesP}).

From Eq.~(\ref{eq:momentummap}) one can see that the momentum map is also
covariantly preserved by the Killing vectors

\begin{equation}
\mathbb{L}_{I}\vec{P}_{J} = 0\, .
\end{equation}

There is still one more consistency check on the momentum map: the
quaternionic K\"ahler two-form is $SU(2)$-covariantly closed. To ensure that
this property is consistent with Eq.~(\ref{eq:triholomorphic}) we must check
that the covariant Lie derivative commutes with the $SU(2)$-covariant exterior
derivative, in analogy to the commutation between standard Lie derivatives and
exterior derivatives. This requirement leads us to the condition

\begin{equation}
\mathcal{L}_{I}\vec{\omega}-d\vec{\eta}_{I}
-2\vec{\omega}\times\vec{\eta}_{I}=0\,.
\label{eq:bomegarelation}
\end{equation}

Notice that this relation between the two $SU(2)$ connections is in principle
independent of Eq.~(\ref{eq:bomegaP}). After substitution of
Eq.~(\ref{eq:bomegaP}) in Eq.~(\ref{eq:bomegarelation}) the latter becomes the
differential definition of $\vec{P}_{I}$, Eq.~(\ref{eq:defmomentummap}).

Eq.~(\ref{eq:defmomentummap}) can alternatively be used to solve the
Killing vectors in terms of the derivatives of the momentum map,

\begin{equation}
k_{I}{}^{X} 
= 
{\textstyle\frac{2}{3}}\vec{J}^{XY}\cdot\mathfrak{D}_{Y}\vec{P}_{I}\, .
\end{equation}

In view of this relation $\vec{P}_{I}$ is sometimes called the prepotential.

The moment map assigns a triplet of real numbers to each Killing vector. The
Killing vectors realize the algebra of $G$. Eq.~(\ref{eq:PtimesP}) can also be
understood as a realization of the algebra of $G$ in terms of $\vec{P}_{I}$,
$\vec{J}_{XY}$ being the symplectic structure used to define the Poisson
brackets which are the left hand side of Eq.~(\ref{eq:PtimesP}).

In summary, given a Killing vector of the metric $g_{XY}(q)$ we can always
construct the momentum map $\vec{P}_{I}$ by Eq.~(\ref{eq:momentummap}). Next
we define the covariant Lie derivative along the Killing vector by means of
the connection $\vec{\eta}_{I}$.  This covariant Lie derivative enjoys the
algebraic and differential properties of a pure Lie derivative and also
commutes with covariant exterior derivatives. The Killing vector becomes
automatically covariantly tri-holomorphic according to
Eq.~(\ref{eq:triholomorphic}).


\subsection{$SU(2)$ transformations induced by $G$}

Let us now consider the momentum map as a composite spacetime field over which
depends only on the $q^{X}$s. Under general variations $\delta q^{X}$
%
%
and using the definition of the momentum map~(\ref{eq:defmomentummap}),

\begin{equation}
\delta\vec{P}_{I} 
=
-\delta q^{X}\left(
{\textstyle\frac{1}{2}}\vec{J}_{XY}k_{I}{}^{Y}
+2\vec{\omega}_{X}\times\vec{P}_{I}\right)\, .
\end{equation}

\noindent
If this transformation is a $G$-gauge transformation $\delta_{\Lambda} q^{X} =
-g\Lambda^{J}k_{J}{}^{X}$, taking into account Eq.~(\ref{eq:PtimesP}), we
obtain

\begin{equation}
\delta_{\Lambda} \vec{P}_{I} 
= 
-gf_{IJ}{}^{K} \Lambda^{J} \vec{P}_{K}
+2g\Lambda^{J} \vec{\eta}_{J}\times\vec{P}_{I}\, ,
\end{equation}

\noindent
which is the adjoint action of $G$ on $\vec{P}_{I}$ plus an induced $SU(2)$
gauge transformation with parameter $-g\Lambda^{J}\vec{\eta}_{J}$ which is
present even if $G$ is Abelian. This is the mechanism through which $G$ can
act on objects such as the spinors of the supergravity theory which only have
$SU(2)$ indices, opening the doors to the gauging of groups larger than
$SU(2)$: if the gravitino transforms under standard $SU(2)$ transformations
according to

\begin{equation}
\delta_{\lambda}\psi_{\mu}^{i} = 
i\psi_{\mu}^{j} \vec{\sigma}_{j}{}^{i} \cdot \vec{\lambda}\, ,
\end{equation}

\noindent
where $\vec{\lambda}$ is the infinitesimal $SU(2)$ parameter, then, under
$G$-gauge transformations it will undergo a similar transformation with
$\vec{\lambda} = -g\Lambda^{I}\vec{\eta}_{I}$.

Thus, in $G$-gauged supergravity the pullback of the $SU(2)$ connection 
that couples to the spinors of the theory has to be replaced by 

\begin{equation}
\vec{B} 
\equiv
\vec{A} +{\textstyle\frac{1}{2}} gA^{I} \vec{P}_{I}\, ,
\hspace*{10mm}
\vec{A} \equiv dq^{X} \vec{\omega}_{X}\, ,
\end{equation}

\noindent
to take into account the $SU(2)$ transformations induced by $G$-gauge
transformations, which act on it as

\begin{equation}
\delta_{\Lambda}\vec{B}
=
-2(-g\Lambda^{I}\vec{\eta})\times\vec{B}
+d(-g\Lambda^{I}\vec{\eta})\,.
\end{equation}

\noindent
The covariant derivative on these objects is

\begin{equation}
\mathfrak{D}_{\mu}\psi_{\nu}^{i} 
= 
\nabla_{\mu}\psi_{\nu}^{i}  +\psi^{j}B_{\mu j}{}^{i}\, .
\end{equation}


\end{document}